\documentclass[pra,aps,a4paper,twocolumn,amsmath,amssymb,floatfix,nobalancelastpage,superscriptaddress,notitlepage,accepted=2021-09-13]{quantumarticle}
\pdfoutput=1
\usepackage{graphicx}
\usepackage{dcolumn}
\usepackage{bm}
\usepackage[dvipsnames]{xcolor}
\usepackage[normalem]{ulem}
\usepackage[makeroom]{cancel}
\usepackage[colorlinks=true,citecolor=cyan]{hyperref}

\newcommand{\beq}{\begin{equation}}
\newcommand{\eeq}{\end{equation}}

\begin{document}

\preprint{}
\title{Determining quantum phase diagrams of topological Kitaev-inspired models on NISQ quantum hardware}

\author{Xiao Xiao}
\affiliation{Department of Physics, North Carolina State University, Raleigh, North Carolina 27695, USA}

\author{J.~K.~Freericks}
\affiliation{Department of Physics, Georgetown University, 37th and O Sts. NW, Washington, DC 20057 USA}

\author{A.~F.~Kemper}
\email{akemper@ncsu.edu}
\affiliation{Department of Physics, North Carolina State University, Raleigh, North Carolina 27695, USA}

\begin{abstract}
Topological protection is employed in fault-tolerant error correction and in developing quantum algorithms with topological qubits. But, topological protection \textit{intrinsic to models being simulated}, also robustly protects calculations, even on NISQ hardware. We illustrate how this works by simulating Kitaev-inspired models on IBM quantum computers and accurately determining their phase diagrams. This requires constructing conventional quantum circuits for Majorana braiding to prepare the ground states of Kitaev-inspired models. The entanglement entropy is then measured to calculate the quantum phase boundaries. We show how maintaining particle-hole symmetry when sampling through the Brillouin zone is critical to obtaining high accuracy. This work illustrates how topological protection intrinsic to a quantum model can be employed to perform robust calculations on NISQ hardware, when one measures the appropriate protected quantum properties. It opens the door for further simulation of topological quantum models on quantum hardware available today.

\end{abstract}
\vspace{1.0cm}

\maketitle

\section{introduction}

Quantum computers are fragile and susceptible to rapid decoherence \cite{Schlosshauer_RMP_2005}, and the use of topological protection has been proposed as a potential remedy. Typically, there are two propositions to make quantum computation practical for deep circuits: (i) fault-tolerant quantum computers, which use logical qubits and error correcting codes (like the surface code) to correct all errors that appear during computation \cite{Steane_PRL_1996,Knill_arXiv_1996,Aharonov_ACM_1997,Terhal_RMP_2015,Aliferis_QIC_2005, Bacon_PRA_2006,Fowler_PRA_2012,Tuckett_PRL_2018,Chao_prl_2018,Li_PRX_2019} and (ii) topological quantum computers, which use topological qubits, naturally protected from the environment, to perform intrinsically error-free computations \cite{Kitaev_AP_1997,Nayak_RMP_2008,Sarma_NPJ_2015}. Here, we propose a third avenue restricted to the simulation of specific models: when the model under study has non-trivial topological protections, the model's topological properties (rather than the hardware's) can be leveraged to identify its sharp phase transitions as the parameters are tuned.

A paradigmatic example of a ground state with non-trivial topological properties, occurs in the Kitaev spin model on the honeycomb lattice \cite{Kitaev_AP_2006}, which has spawned a class of related models with similar properties \cite{Yao_PRL_2007,Yang_PRB_2007,Feng_PRL_2007,Lee_PRL_2007,Si_arXiv_2007,Yu_EPL_2008,Baskaran_arXiv_2009,Mandal_PRB_2009,Ryu_PRB_2009,Yao_PRL_2009,Wu_PRB_2009,Tikhonov_PRL_2010,Chern_PRB_2010,Wang_PRB_2010,Lahtinen_PRB_2010,Kells_NJP_2011,Yao_PRL_2011, Lai_PRB_2011,Chua_PRB_2011,Nakai_PRB_2012,Nussinov_RMP_2015,Hermanns_PRL_2015,O'Brien_PRB_2016,Chen_PRL_2018,Miao_arXiv_2018,Miao_PRB_2018}.
Their solution involves the braiding and fusing of anyons (in this case Majorana fermions), just as algorithms employing topological qubits do. Ideally, topological quantum computers could be used to solve models of this type. Unfortunately, it has proved to be extremely difficult to create topological qubits in quantum hardware \cite{Stern_Science_2013}.

Instead, we illustrate that these Kitaev-inspired models can be solved on conventional quantum computers with the braiding operations encoded in low-depth conventional quantum gates.  Moreover, we can utilize the distinct topological properties of ground states to clearly distinguish the phase transitions in this class of models for system sizes where noise makes this difficult if one uses a local quantity instead. These topological quantum simulations are carried out on IBM's transmon-based ``conventional'' quantum computing hardware. 

Typically, determining the phase boundaries in these models is challenging because they lack a local order parameter. The standard approach is to calculate the entanglement entropy~\cite{Yao_PRL_2010,Meichanetzidis_PRB_2016}, but this requires a large system size, beyond what is currently available in NISQ machines. Instead, we develop and use particle-hole symmetry-enforced circuits encoding twisted boundary conditions, enabling determination of ground-state properties through the entire Brillouin zone with only a modest number of qubits. Due to the discrete topological nature of the quantum phase transition, these circuits can identify the different phases by revealing discontinuities in the observables, even if the jump at the discontinuity is diminished due to noise.

The paper is organized as follows. We begin by discussing the general procedure for diagonalizing Kitaev-inspired models
in Sec.~\ref{sec:exact_solution}. This is followed by a step-by-step translation of the procedure to a sequence of quantum
circuits in Sec.~\ref{sec:circuit_construction}, with a particular focus on the braiding operations. We finish the
section with a discussion on the generalization and scaling of the quantum circuits in Sec.~\ref{sec:scaling}.
With the circuits from Sec.~\ref{sec:circuit_construction} in hand, we apply these to the 1D Kitaev spin chain and the Kitaev honeycomb model in Sec.~\ref{entanglement_entropy_density_matrix} and measure the entanglement
entropy.
In Sec.~\ref{sec:symmetry_enforced_circuit}, we propose symmetry-enforced circuits to sample the Brillouin zone while maintaining a fixed number of qubits, and demonstrate the application of these circuits by measuring the entanglement spectra of 1D Kitaev spin chain via correlation functions. This enables the determination of the phase diagrams for the 1D Kitaev spin chain and Kitaev honeycomb models, as shown in Sec.~\ref{sec:phasediagram}.  We conclude in Sec.~\ref{sec:conclusion}. 

\begin{figure*}[htpb]
  \centering
  \includegraphics[width=1.4\columnwidth]{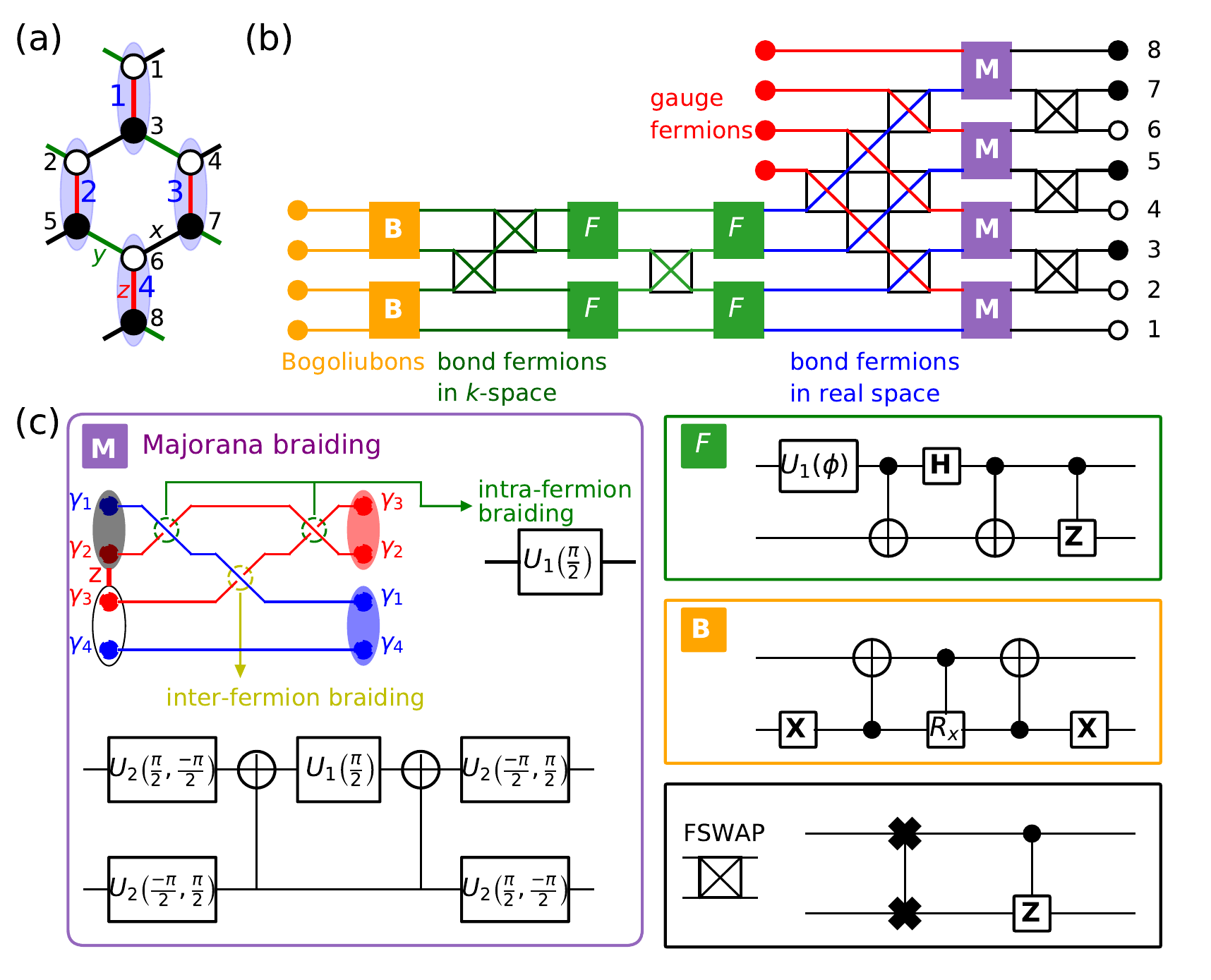}
  \caption{Quantum circuit for the Kitaev honeycomb model with braiding: (a) The lattice structure of the Kitaev honeycomb model with the numbers denoting the ordering along the Jordan-Wigner string and the blue shaded regions indicating four unit cells; (b) the quantum circuit used to prepare the ground-state wave function (each qubit is initially in the $|0\rangle$ state); (c) details of the elements of the quantum circuit shown in (b). In (a) the $x$-, $y$- and $z$-bonds are shown in black, green and red respectively. The operation $F$ shown in (c) stands for Fourier transformation operations in (b), the operation $B$ in (c) denotes Bogoliubov transformation operations in (b), and the FSWAP operation in (c) stands for the operation exchanging the order of the neighboring qubits representing fermions in (b).}
   \label{fig1}
\end{figure*}

\section{Kitaev-inspired models and the quantum circuit to prepare their ground states} \label{kitaev_circuit}

\subsection{Kitaev-inspired models and exact solvability} \label{sec:exact_solution}

Based on the seminal work by Kitaev \cite{Kitaev_AP_2006}, significant effort was made to generalize the exact solution procedures to other strongly correlated spin models \cite{Yao_PRL_2007,Yang_PRB_2007,Feng_PRL_2007,Lee_PRL_2007,Si_arXiv_2007,Yu_EPL_2008,Baskaran_arXiv_2009,Mandal_PRB_2009,Ryu_PRB_2009,Yao_PRL_2009,Wu_PRB_2009,Tikhonov_PRL_2010,Chern_PRB_2010,Wang_PRB_2010,Lahtinen_PRB_2010,Kells_NJP_2011,Yao_PRL_2011,Lai_PRB_2011, Chua_PRB_2011,Nakai_PRB_2012,Nussinov_RMP_2015,Hermanns_PRL_2015,O'Brien_PRB_2016,Chen_PRL_2018,Miao_arXiv_2018,Miao_PRB_2018}. These models can be solved exactly by using the same strategy proposed by Kitaev \cite{Kitaev_AP_2006} and as such we refer to them as ``Kitaev-inspired'' models in this work. 

The exact solution of these models begins with assigning a proper Jordan-Wigner (JW) string ordering for all the sites, along which the JW transformation is applied. After the JW transformation, the coupling terms involving transverse ($x$ and $y$) components of spins can be transformed to be quadratic in terms of Majorana fermions
(denoted here by $\gamma$ operators). 

The coupling terms between the $z$-component of spins are transformed into density-density interactions between the two sites. In terms of Majorana fermions, the occupation number at a generic site $j$ is proportional to $\gamma_{j} \eta_{j}$, so that the $z$-component coupling term between sites $i$ and $j$ is $\propto \gamma_{i} \eta_{i} \gamma_{j} \eta_{j}$. 
Here, the fermion at site $i$ is composed of Majorana fermions $\eta_i$ and $\gamma_i$. 
According to Kitaev \cite{Kitaev_AP_2006,Yao_PRL_2010}, and in agreement with subsequent numerical verification \cite{Chen_PRL_2018} for the ground state in the thermodynamic limit, the product of the two $\eta$ Majorana fermions reduces to a site-independent gauge field due to local conservations \cite{Yao_PRL_2010,Schmoll_PRB_2017}. Consequently, as long as we are interested in ground states of the model, the $z$-component coupling terms can be treated as if they are only quadratic (in terms of $\gamma$ Majorana fermions) and the quadratic in the $\eta$ fermions can be replaced by the constant gauge field value. 

To complete the exact solution, new fermion operators are defined from the $\gamma$ Majorana fermions, and the resulting models can be further diagonalized by Fourier transformations followed with Bogoliubov transformations. In the following sections, we will denote the composite fermion constructed from two $\eta$ Majorana fermions as the \textit{gauge fermion}, and the composite fermion constructed from the two $\gamma$ Majorana fermions as the \textit{bond fermion}.

To explicitly demonstrate how to study these models on quantum hardware, we focus on the original Kitaev honeycomb model and its $1D$ simplification, the Kitaev spin chain (see Appendix \ref{appendix a} for more details), in the main text. Because of the generality of the exact solution procedures for the Kitaev-inspired models, the methods presented here can be readily applied to other Kitaev-inspired models.
These include, for example, the generalizations of the original Kitaev honeycomb model to different lattices \cite{Yao_PRL_2007,Yang_PRB_2007,Feng_PRL_2007,Lee_PRL_2007,Tikhonov_PRL_2010,Chua_PRB_2011}, different dimensions \cite{Si_arXiv_2007,Mandal_PRB_2009,Ryu_PRB_2009,O'Brien_PRB_2016,Chen_PRL_2018,Miao_arXiv_2018}, and higher spins \cite{Wu_PRB_2009,Wang_PRB_2010}. In the Appendix \ref{appendix c}, one additional example of applying our approach to the $1D$ BCS-Hubbard model is provided. 

\subsection{Quantum circuit constructions}
\label{sec:circuit_construction}

To illustrate how to construct the quantum circuit that prepares the ground states of Kitaev-inspired models, we focus on the original Kitaev model constructed on the honeycomb lattice with 8 sites (resulting in $2\times 2$ unit cells) as shown in Fig.~\ref{fig1}(a). 
To prepare the ground-state wave function in real space, we begin in momentum space (after all required transformations), where the Hamiltonian is diagonalized (and then go backwards through all of the steps required for solving the problem to end up back in real space). The ground state (in momentum space) is then written as a simple product state. In particular, for the model under consideration, the ground state is just the vacuum state $|0000\rangle$, with the qubits representing the four momentum points of the elementary excitations with (from left to right) $(k_x,k_y) = (-\pi/2,-\pi/2)$,  $(k_x,k_y) = (\pi/2,\pi/2)$, $(k_x,k_y) = (\pi/2,-\pi/2)$ and $(k_x,k_y) = (-\pi/2,\pi/2)$, respectively (We work with periodic boundary conditions). 
In the quantum circuit shown in Fig.~\ref{fig1}(b), these momentum points correspond to the qubits indicated by the orange dots on the far left, ordered from bottom to top. 
The circuit construction for the Bogoliubov and fermionic Fourier transform (shown in Fig.~\ref{fig1}(c)) are detailed in Refs.~\cite{Verstraete_PRA_2009,Lierta_Qu_2018}, which we reproduce in the Appendix \ref{circuit_Bogoliubov_Fourier} for completeness.

Here we divide the $8$ sites into four unit cells as shown by the blue shaded regions in Fig.~\ref{fig1}(a) so that each unit cell contains one gauge fermion and one bond fermion. In particular, we define the unit cell $1$ containing sites $1$ and $3$, the unit cell $2$ containing sites $2$ and $5$, the unit cell $3$ containing sites $4$ and $7$, and the unit cell $4$ containing sites $6$ and $8$. Then the $4$ qubits after the Fourier transformations from bottom to top in Fig.~\ref{fig1}(b) correspond to the bond fermions in the unit cells from $1$ to $4$. The gauge fermions in the unit cells from $1$ to $4$ are represented by the $4$ red dots from bottom to top in Fig.~\ref{fig1}(b). 

Corresponding to the exact solution procedure that simplifies the $z$-component coupling terms, only the braiding of the Majorana fermions needs to be performed in the quantum circuit. To facilitate the implementation of the braiding operations, the bond fermion and the gauge fermion belonging to the same unit cell must be neighbors. 
In the quantum circuit, this is achieved by a series of fermion swap operations after the Fourier transformations as shown in Fig.~\ref{fig1}(b). Then, the braiding operations are implemented on the $4$ sets of neighboring qubits as shown by the purple boxes in Fig.~\ref{fig1}(b). The braiding circuit is discussed in detail below. The final step is to rearrange the fermions according to the JW string ordering, and this is done by the $3$ swap operations shown in Fig.~\ref{fig1}(b).

\subsubsection{Quantum Circuits for Braiding Operations}

The strategy we use to create quantum circuits is to recognize that the braiding operations can be thought of as an intermediary that maps the conventional fermions to themselves---after identifying this mapping, we can then construct circuits within the conventional fermion language. We describe how this is done next. 

The simplest braiding is the braiding of  two Majorana fermions to construct a conventional fermion. A conventional fermion, which can be written as $f=(\gamma_2 + i\gamma_1)/2$, is connected to another conventional fermion defined by $\tilde{f}=(\gamma_1 + i\gamma_2)/2$ via a braiding operation on the Majorana fermions. Recall that the braiding operator (in terms of Majorana fermions) is given by \cite{Ivanov_PRL_2001,Schmoll_PRB_2017}
\beq
B_{i,j}^{\pm} = \frac{1}{\sqrt{2}} \left( 1 \pm \gamma_{i} \gamma_j \right).
\eeq
In this first example, the two Majorana fermions come from the same conventional fermion. After expressing the Majorana fermions in terms of the conventional fermions, we find that
\begin{align}
B_{1,2}^{\pm} &= \frac{1}{\sqrt{2}} \left[ 1 \pm i(f^\dag-f) ( f+f^\dag ) \right] \nonumber \\
&= \frac{1}{\sqrt{2}} \left[ 1 \pm i( f^\dag f - ff^\dag ) \right],
\end{align}
where $ff$ and $f^\dag f^\dag$ vanish due to the Pauli principle. To express this operator as a matrix in the qubit representation, we map the computational basis $\left( |0\rangle, |1\rangle \right)$ onto the conventional fermion number basis yielding
\beq
B_{1,2}^{\pm} = B_\text{in}^{\pm} = \frac{1}{\sqrt{2}} \left(\begin{array}{cc} 1\mp i & 0 \\ 0 & 1\pm i \end{array}\right).
\eeq

In our second example, we consider the case of the braiding of two Majorana fermions that belong to two different conventional fermions. The conventional fermions are
\beq
\begin{cases}
f_1 = (\gamma_2 + i\gamma_1)/2, \\
f_2 = (\gamma_4 + i\gamma_3)/2,
\end{cases}
\eeq
when expressed in terms of the different Majorana fermions.

There are four different ways to braid these Majorana fermions. Because braiding operations do not commute, we have to carefully specify the ordering scheme for each set of fermions. We choose an ascending order for the conventional fermions and also for the Majorana fermions within the same conventional fermion. With this specification, we find that the different  inter-fermion braidings can be expressed as the product of the braidings within the same fermions and one more braiding, denoted by $B_{2,3}^{\pm}$. This is the fundamental braiding between two conventional fermions and is denoted $B_\text{ex}$ in the later discussions. Replacing the Majorana operators by fermion operators, we find that
\begin{align}
B_\text{ex}^{\pm} &= \frac{1}{\sqrt{2}} \left( 1 \pm i(f_1+f_1^\dag) (f_2^\dag-f_2) \right), \nonumber \\
&=\frac{1}{\sqrt{2}} \left( 1 \pm i(f_1^\dag f_2^\dag - f_1^\dag f_2 + f_1 f_2^\dag - f_1 f_2) \right).
\end{align}
Again, we using a conventional number basis $\left( |0_{f_1} 0_{f_2}\rangle, |1_{f_1} 0_{f_2}\rangle, |0_{f_1} 1_{f_2}\rangle, |1_{f_1} 1_{f_2}\rangle \right)$, yields a matrix representation for the inter-fermion braiding
\beq
B_\text{ex}^{\pm} = \frac{1}{\sqrt{2}} \left(\begin{array}{cccc} 1 & 0 & 0 & \pm i \\ 0 & 1 & \mp i & 0 \\ 0 & \mp i & 1 & 0 \\ \pm i & 0 & 0 & 1 \end{array}\right).
\eeq
Next, we decompose this braiding operation in terms of conventional quantum gates. Since the clockwise braiding is the Hermitian conjugate of the anti-clockwise braiding, we focus on the clockwise braiding. We first consider the braiding within the same fermion, and we notice that
\beq
B_\text{in}^{+} = \frac{1}{\sqrt{2}} \left(\begin{array}{cc} 1- i & 0 \\ 0 & 1+ i \end{array}\right) = e^{-i\pi/4} \left(\begin{array}{cc} 1 & 0 \\ 0 & i \end{array}\right),
\eeq
so $B_\text{in}$ can be realized by the phase gate $U_1(\pi/2)$ as shown in the left panel in Fig.~\ref{fig1}(c). Similarly, we find:
\beq
B_\text{ex}^{+} = e^{-i\frac{\pi}{4} \sigma_y \otimes \sigma_y}.
\eeq
So, the quantum circuit can be immediately constructed via the general scheme proposed by Vidal \textit{et al.} \cite{Vidal_PRA_2004}. The detailed circuit for $B_\text{ex}^{+}$ is shown in the left panel in Fig.~\ref{fig1}(c).

For the Kitaev-inspired models, the braiding operations are used to change the positions of Majorana fermions, allowing Majorana fermions belonging to two different conventional fermions in one unit cell to recombine into new fermions. This implies that all braiding processes can be constructed with one inter-fermion braiding and a few intra-fermion braidings. The inter-fermion braiding  changes the original two-qubit states into a superposition of their particle-hole partners with a $\pm\pi/2$ phase difference. Intra-fermion braidings behave differently depending on whether they are performed before or after the inter-fermion braiding. If the intra-fermion braiding is performed before the inter-fermion braiding, it introduces a $\pm \pi/4$ phase to the fermion. However, if the intra-fermion braiding is performed after the inter-fermion braiding, it introduces a superposition of the original state (before inter-fermion braiding) and its particle-hole partner (with a $\pm\pi/2$ phase difference). These basic properties of braiding help us construct braiding circuits based on local constraints.

The first constraint imposed on the braiding operations comes from the local $z$-bonds. For Kitaev-inspired models, the local fermionic Hamiltonian coupling between two sites by a $z$-bond is
\beq
H_z = J_z \left( 2c_{1}^\dag c_{1} - 1 \right) \left( 2c_{2}^\dag c_{2} - 1 \right).
\eeq
where $c_{1}$ and $c_{2}$ are the annihilation operators for the fermions on the site $1$ and site $2$. The two-qubit states are denoted by $|n_{1} n_{2}\rangle$. By the braiding operation $\mathcal{U}_B$, this Hamiltonian can be re-expressed in terms of bond and gauge fermions as
\beq \label{interaction_term}
\tilde{H}_z = \mathcal{U}_B^\dag H_z \mathcal{U}_B = J_z D \left( 2f^\dag f - 1 \right),
\eeq
where $D$ is the local gauge field and $f$ stands for the bond fermion. Correspondingly, after the braiding transformation, the two-qubit states here are $|n_f n_g\rangle$ with $n_g$ ($n_f$) denoting the occupation of the gauge (bond) fermion. This implies that the states $|n_f n_g\rangle$ are transformed to the states $|n_1 n_2\rangle$ as follows:
\beq \label{braiding}
\begin{cases}
\mathcal{U}_B |0_f0_g\rangle = \alpha_{0_f0_g}^{0_10_2} |0_10_2\rangle + \alpha_{0_f0_g}^{1_11_2} |1_11_2\rangle, \\
\mathcal{U}_B |1_f1_g\rangle = \alpha_{1_f1_g}^{0_10_2} |0_10_2\rangle + \alpha_{1_f1_g}^{1_11_2} |1_11_2\rangle, \\
\mathcal{U}_B |1_f0_g\rangle = \alpha_{1_f0_g}^{1_10_2} |1_10_2\rangle + \alpha_{1_f0_g}^{0_11_2} |0_11_2\rangle, \\
\mathcal{U}_B |0_f1_g\rangle = \alpha_{0_f1_g}^{1_10_2} |1_10_2\rangle + \alpha_{0_f1_g}^{0_11_2} |0_11_2\rangle,
\end{cases}
\eeq
where $\alpha_{n_1 n_2}^{n_f n_g}$ denotes the transformation coefficient. The two Hamiltonians and their corresponding states are connected by braiding operations, so the matrix elements of the Hamiltonian (before and after transformation) needs to be consistent. A direct calculation yields
\begin{align}
&\langle 0_f0_g| \tilde{H}_z |0_f0_g\rangle = -\langle 1_f1_g| \tilde{H}_z |1_f1_g\rangle \nonumber \\
= &-\langle 1_f0_g| \tilde{H}_z |1_f0_g\rangle = \langle 0_f1_g| \tilde{H}_z |0_f1_g\rangle = -J_z D,
\end{align}
and
\begin{align}
&-\langle 0_f0_g| \mathcal{U}_B^\dag H_z \mathcal{U}_B |0_f0_g\rangle = -\langle 1_f1_g| \mathcal{U}_B^\dag H_z \mathcal{U}_B |1_f1_g\rangle \nonumber \\
= &\langle 1_f0_g| \mathcal{U}_B^\dag H_z \mathcal{U}_B |1_f0_g\rangle = \langle  0_f1_g| \mathcal{U}_B^\dag H_z \mathcal{U}_B |0_f1_g\rangle = -J_z.
\end{align}
Therefore, the consistency imposes the constraint $D=1$ for $n_g=1$ and $D=-1$ for $n_g=0$.

This local constraint imposed by $z$-bonds is general for Kitaev-inspired models. But, the local constraints imposed by other coupling terms (\textit{i.e.} $x$-, and $y$-bonds for the models considered in this work) depend on the JW strings, because those strings determine the ordering of bond fermions emerging from different braiding operations. By using the same analysis as shown here for the $z$-bond, the details of the braiding operations can be further determined from the constraints set by other coupling terms. For the Kitaev honeycomb model on the honeycomb lattice with the JW string in the ordering given in Fig.~\ref{fig1}(a), the braiding operation can be implemented by the circuit shown in the left panel in Fig.~\ref{fig1}(c). The details of constructing the braiding circuit for the $1D$ Kitaev spin chain can be found in Appendix \ref{appendix b}.

\subsection{Generalization and scaling}
\label{sec:scaling}

We have illustrated how to construct the quantum circuit to prepare the ground state of the Kitaev honeycomb model with $2\times2$ unit cells in real space by reversing the methodology for the exact solution of the model. Because the exact solution procedures can be generally applied to a large class of Kitaev inspired models \cite{Yao_PRL_2007,Yang_PRB_2007,Feng_PRL_2007,Lee_PRL_2007,Si_arXiv_2007,Yu_EPL_2008,Baskaran_arXiv_2009,Mandal_PRB_2009,Ryu_PRB_2009,Yao_PRL_2009,Wu_PRB_2009,Tikhonov_PRL_2010,Chern_PRB_2010,Wang_PRB_2010,Lahtinen_PRB_2010,Kells_NJP_2011,Yao_PRL_2011,Lai_PRB_2011, Chua_PRB_2011,Nakai_PRB_2012,Nussinov_RMP_2015,Hermanns_PRL_2015,O'Brien_PRB_2016,Chen_PRL_2018,Miao_arXiv_2018,Miao_PRB_2018}, the circuit construction method illustrated here can be readily extended for other Kitaev-inspired models. The quantum circuits represented by the orange and green boxes in Fig.~\ref{fig1} are elementary components for the implementation of generic Bogoliubov and Fourier transformations. Though the braiding circuits shown in Fig.~\ref{fig1} would change according to different definitions of JW strings for the different models, the methodology to construct circuit will be the same. For example, in Appendix \ref{appendix c} we show that the circuit construction method can be applied to the $1D$ BCS-Hubbard model \cite{Chen_PRL_2018}, which is another example of a Kitaev-inspired model.

Before we close this section, we would like to discuss the scaling of the circuit for the Kitaev honeycomb model. Because of the Fourier transformation, the system size must contain $\mathcal{N} = 2^n\times 2^m$ unit cells with $2^n$ along the $x$ direction and $2^m$ along the $y$ direction. To prepare the ground state of such a system, the required number of qubits is $2\mathcal{N}$. The Bogliubov transformation is implemented on  opposite momentum point pairs, so we require $\mathcal{N}/2$ two-qubit Bogoliubov transformations (the corresponding circuit shown in Fig.~\ref{fig1}(c)) in building the full circuit. We know that for $N$ momentum points along one direction, the Fourier transformation would require $N \log N$ two-qubit Fourier transformations (the circuit as shown in Fig.~\ref{fig1}(c)) \cite{Verstraete_PRA_2009,Lierta_Qu_2018,Ferris_PRL_2014}. To complete the Fourier transformation along both the $x$ and $y$ directions, the total number of two-qubit Fourier transformations is still $\mathcal{N} \log \mathcal{N}$. The braiding operations are performed within each unit cell, so $\mathcal{N}$ two-qubit braiding operations shown in Fig.~\ref{fig1}(c) are required. Finally, we count the number of swap boxes (shown in Fig.~\ref{fig1}(c)) in the quantum circuit, and they would appear in the following four places according to the generic quantum circuit shown in Fig.~1(b): (i) after the Bogoliubov transformation to facilitate the Fourier transformation along the $y$ direction; (ii) after the transformation in the $y$ direction to facilitate the Fourier transformation along the $x$ direction; (iii) when pairing the bond fermion with the gauge fermion in the same unit cell; and (iv) when reordering the fermions to map to the original spin representation. The number of swap boxes is then $(\mathcal{N}/4+1)(\mathcal{N}/2-1)$ in (i), $\mathcal{N}/4$ in (ii), and $(\mathcal{N}^2-\mathcal{N})/2$ in (iii). In the implementation of quantum circuit the swap boxes in (iv) can be saved by relabeling the qubits and counting the sign changes, which can be done classically. In summary, in the quantum circuit to prepare the ground state of the Kitaev honeycomb model, the number of two-qubit Bogoliubov transformation circuits and the number of two-qubit braiding circuits are each proportional to the lattice size as $ \propto \mathcal{O}(\mathcal{N})$. The number of two-qubit Fourier transformation circuits is $\mathcal{O}(\mathcal{N}\log\mathcal{N})$, and the number of swap boxes is proportional to the square of the lattice size as $ \propto \mathcal{O}(\mathcal{N}^2)$. A similar analysis can be done for other Kitaev-inspired models as well, but we do not do so here.

\section{Measuring the ground state properties of Kitaev-inspired models on real quantum hardware} \label{entanglement_entropy_density_matrix}

\begin{figure}[h]
  \centering
  \includegraphics[width=1\columnwidth]{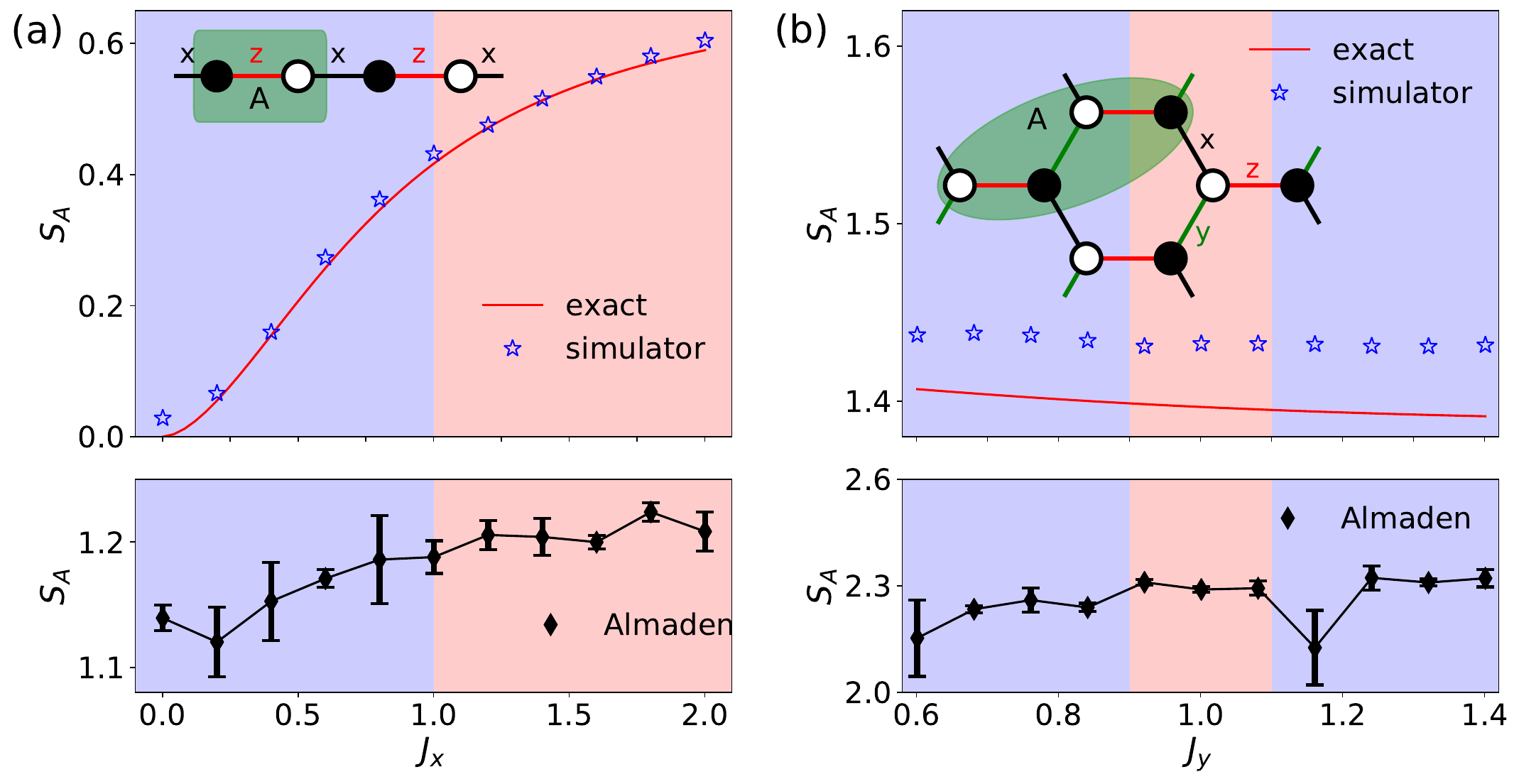}
  \caption{ Quantum simulation of the entanglement entropy of Kitaev-inspired models: (a) the $1D$ Kitaev spin chain with $J_z=1$ and varying $J_x$; (b) the $2D$ Kitaev spin model on the honeycomb lattice with $J_z=1$, $J_x=0.1$ and varying $J_y$. In both (a) and (b), the lattice structure and the definition of the subsystem `A' (the green shaded parts) are shown in the insets of the upper panels, the comparison of the exact results and the results from the simulator are shown in the upper panel, and the results from IBMQ-Almaden are shown in the lower panel. Each data point from IBMQ-Almaden were obtained by performing five independent experiments, and for each of them $N=8196$ shots were used. The simulator data for the $1D$ Kitaev spin chain was obtained with $N=8196$ shots, while the simulator data for the $2D$ Kitaev spin model was obtained with $N=81960$ shots.}
   \label{fig2}
\end{figure}

By using the procedures described in the section \ref{kitaev_circuit}, we can construct the quantum circuits to prepare the ground states for the $1D$ Kitaev spin chain containing two unit cells (including $4$ sites overall) and the Kitaev honeycomb model containing $2\times 2$ unit cells (including $8$ sites overall). The lattice configuration of the two models are shown in the the insets in Fig.~\ref{fig2}(a) and (b) respectively. Indeed, these circuits can correctly prepare the ground states of the two models as shown in Appendix \ref{appendix d}. Given that the ground states of the models are topological, a useful quantity to characterize them is the entanglement entropy. To measure the entanglement entropy, the lattice of the model is divided into two subsystems; the entanglement entropy of the subsystem $A$ (the green colored regions in the insets of Fig.~\ref{fig2}) is found by projectively measuring the density matrix with the help of the maximum likelihood estimator \cite{Smolin_PRL_2012}. While this approach clearly does not scale, it is feasible for these proof-of-principle size systems.

In the upper panel in Fig.~\ref{fig2}(a) we plot the entanglement entropy for the $1D$ Kitaev spin chain as a function of the coupling strength $J_x$ by setting the other coupling strength $J_z=1$. We found that the results obtained from the quantum simulator provided by QISKIT \cite{qiskit2019} differs from the exact results only by statistical errors. This indicates that the quantum circuit can indeed correctly prepare the ground state of the model. Both the exact and simulator results show that the entanglement entropy for the $1D$ Kitaev spin chain increases with  $J_x$, and the derivative is largest at intermediate values of $J_x$ (around $J_x=0.5$). As a comparison, the entanglement entropy obtained from the  IBMQ-Almaden quantum computer is shown in the lower panel in Fig.~\ref{fig2}(a). The overall features of the entanglement entropy versus $J_x$ are captured in the real machine data, \textit{i.~e.} the entanglement entropy increases with $J_x$ with a larger derivative at intermediate values of $J_x$. But the quantitative values deviate significantly due to the noise in the data.

In the upper panel in Fig.~\ref{fig2}(b), the entanglement entropy for the Kitaev honeycomb model is plotted as a function of coupling strength $J_y$ by setting the other two coupling strengths to $J_x=0.1$ and $J_z=1$. The difference between the simulator results and the exact results is due to statistical errors. Because the Hilbert space of the ground state is larger for this model, the statistical errors appear more significant. We found that the entanglement entropy for the model on a $2\times2$ lattice is almost a constant and slightly decreases with increasing $J_y$. The entanglement entropy measured on the machine IBMQ-Almaden is shown in the lower panel of Fig.~\ref{fig2}(b). The real machine data deviates quite a bit from the exact results. Similar to the exact results, the entanglement entropy obtained from the quantum hardware is almost  constant, but it increases slightly with the increasing $J_y$ (unlike the exact results). This larger error is due to the fact that the circuit depth is about three times longer here.

\section{Scanning the Brillouin zone via symmetry-enforced circuits}
\label{sec:symmetry_enforced_circuit}

The ranges of exchange parameters in Fig.~\ref{fig2} cover a regime where both models have quantum phase transitions (in the thermodynamic limit). We observe no signature of these transitions due to finite-size effects (see Appendix \ref{appendix e} for details). To go beyond this limitation of NISQ machines, we employ a grid of shifted momenta to capture the topological phase transition features in the bond fermion sector. With the help of this method, and given the fact that the contribution from the gauge fermion portion is featureless \cite{Yao_PRL_2010}, the entanglement spectra as well as the entanglement entropy can be calculated from the correlation functions in the bond fermion sector only \cite{Peschel_JPA_2003,Vidal_PRL_2003}.

\subsection{Symmetry-enforced circuit}

\begin{figure}[tb]
  \centering
  \includegraphics[width=1\columnwidth]{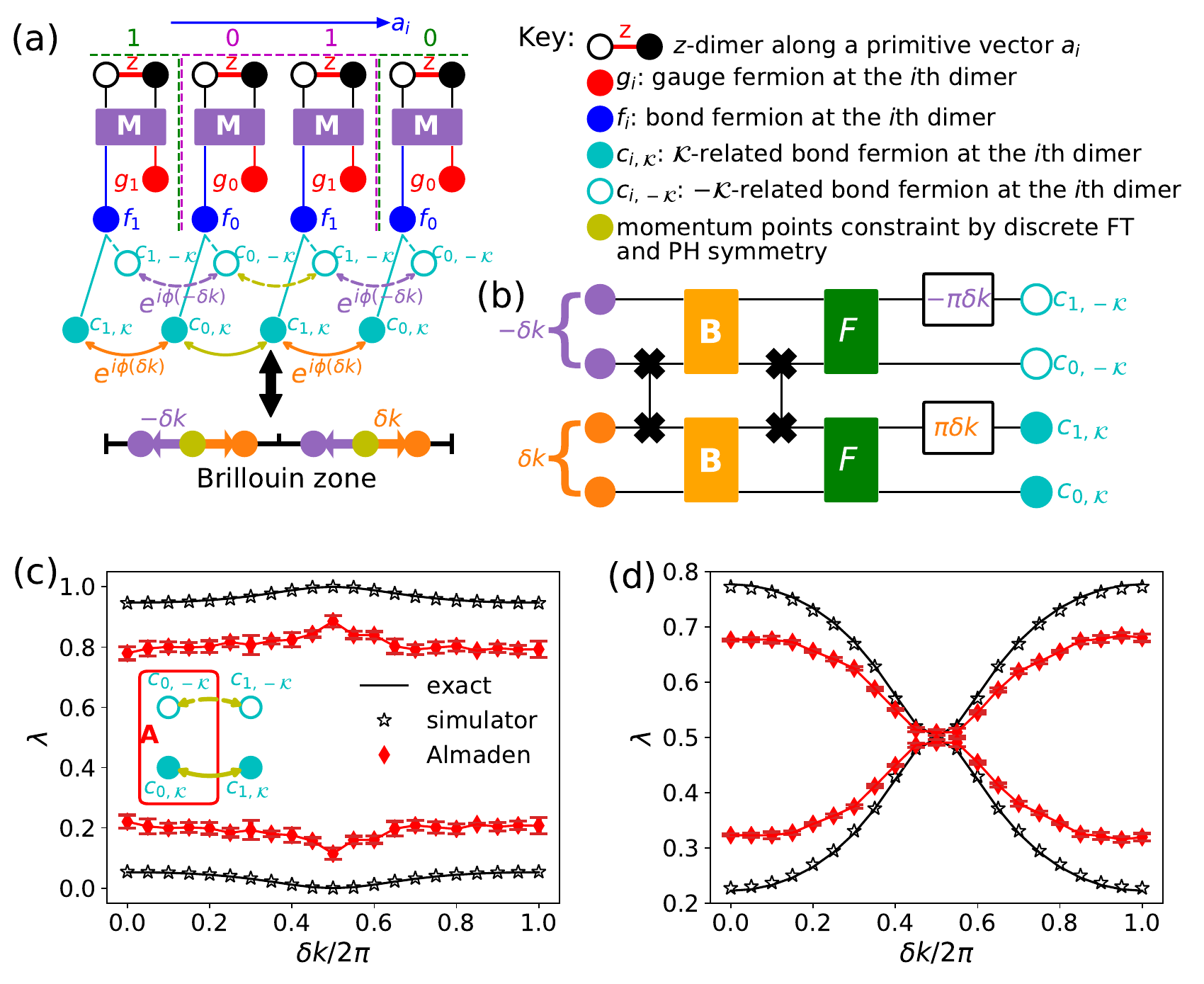}
  \caption{The symmetry-enforced methodology: (a)  symmetry-enforced particle-hole  processes in real and momentum space; (b) the symmetry-enforced circuit; (c) the entanglement spectrum of the gapped phase for the $1D$ Kitaev spin chain with $J_x=0.5<J_z=1$; and (d) the entanglement spectrum of the gapless phase for the $1D$ Kitaev spin chain with $J_x=1.5>J_z=1$.  In (c) and (d), each data point from simulator shown was obtained with $N=8196$ shots, and each data point from IBMQ-Almaden was obtained from three independent experiments, with$N=8196$ shots for each experiment.}
   \label{fig3}
\end{figure}

The Kitaev-inspired models reduce to a block-diagonal form in momentum space after they have been converted to the Majorana Fermion representation. But, because a Bogoliubov transformation is still needed to fully diagonalize the problem, one must be careful to preserve particle-hole (PH) symmetry when constructing the quantum circuit. Consider a system containing $N$ unit cells. The discrete Fourier transformation to an arbitrary set of $N$ momentum points in the BZ, $\mathcal{K} = \{2\pi n/N\ |\ n \in [0,N)\}+\delta k$ can be written as:
\beq \label{shift_momentum}
c_{n,\mathcal{K}} = \frac{1}{\sqrt{N}} \sum_{k\in\mathcal{K}} e^{ikn} c_{k},
\eeq
where $c_{n,\mathcal{K}}$ denotes real-space operators  obtained by Fourier transformation of the set $\mathcal{K}$. Here, $\delta k$ is introduced to shift the momentum points. Physically, it can be achieved by introducing a twist boundary condition as is shown in Fig.~\ref{fig3}(a). In the Fourier transformation circuits, the phase shift requires a phase correction  \cite{Ferris_PRL_2014}. PH symmetry requires that the information from the set with opposition momentum points $-\mathcal{K}$ must also be included. Hence, the symmetry-enforced Fourier transformation is given by:
\beq
c_n = \frac{1}{\sqrt{2}} \left( c_{n,\mathcal{K}} + c_{n,-\mathcal{K}} \right).
\eeq
This indicates that a more complex circuit with twice the number of qubits is required. The demonstration of the symmetry-enforced processes under the particle-hole symmetry for the $1D$ Kitaev spin chain with $N=2$ unit cells is shown schematically in Fig.~\ref{fig3}(a). The corresponding quantum circuit is shown in Fig.~\ref{fig3}(b). Since the Bogoliubov transformation mixes $\pm \mathcal{K}$, the correct momentum set is swapped before the transformation, and swapped back after. In general, the symmetry-enforced circuit can not be simplified, but for special momentum shifts fulfilling the conditions stated in Appendix \ref{appendix f}, a simplification of the symmetry-enforced circuit is possible.

\subsection{Measuring the entanglement of Kitaev-inspired models from correlation functions} \label{correlation}

For Kitaev-inspired models, ground states have uniform gauge configurations, which contribute trivially to the entanglement entropy. The non-trivial quantum information is stored in the Brillouin zone of the bond fermions, whose entanglement properties can be efficiently obtained from the correlation functions.

To begin, the original spin model is mapped to a conventional fermionic model by the Jordan-Wigner transformation followed by a Majorana braiding. In general, the fermionic Hamiltonian can be written as
\beq
H_f = \sum_{i,j} A_{i,j} f_i^\dag f_j + \sum_{i,j} B_{i,j} \left( f_i^\dag f_j^\dag + h.c. \right),
\eeq
where $A_{i,j}$ is the hopping matrix and $B_{i,j}$ is the pairing matrix. The explicit form of the two matrices depends on the details of models. The entanglement entropy of this Hamiltonian is easily obtained by introducing Majorana fermions $f_i = \gamma_{2i-1} + i\gamma_{2i}$ for each physical fermion. In terms of the Majorana fermions, the entanglement entropy of a subsystem of the system is calculated from the eigenvalues of the correlation matrix $\mathcal{C}$ \cite{Peschel_JPA_2003,Vidal_PRL_2003}, whose matrix elements satisfy $C_{m,n} = \frac{1}{2} \text{Tr} \left[\rho \gamma_m \gamma_n\right] = \frac{1}{2} \langle \gamma_m \gamma_n \rangle$ with $\rho$ denoting the density matrix of the system. Then the entanglement entropy is given by:
\beq
S_A = -\frac{1}{2} \sum_{n} [(1-\lambda_n) \ln (1-\lambda_n) + \lambda_n \ln \lambda_n],
\eeq
with $\lambda_n$ the eigenvalues of $\mathcal{C}$. 

To measure the correlation matrix $\mathcal{C}$ on a quantum computer, we note that:
\beq
\begin{cases}
C_{2m-1,2n-1} = \langle f_m^\dag f_n^\dag \rangle + \langle f_m f_n \rangle + \langle f_m f_n^\dag \rangle + \langle f_m^\dag f_n \rangle, \\
C_{2m,2n} = -\langle f_m^\dag f_n^\dag \rangle - \langle f_m f_n \rangle + \langle f_m f_n^\dag \rangle + \langle f_m^\dag f_n \rangle, \\
C_{2m-1,2n} = i\langle f_m^\dag f_n^\dag \rangle - i\langle f_m f_n \rangle + i\langle f_m f_n^\dag \rangle - i\langle f_m^\dag f_n \rangle, \\
C_{2m,2n-1} = i\langle f_m^\dag f_n^\dag \rangle - i\langle f_m f_n \rangle - i\langle f_m f_n^\dag \rangle + i\langle f_m^\dag f_n \rangle.
\end{cases}
\eeq
This suggests that we need to measure four expectation values: $\langle f_m^\dag f_n^\dag \rangle$, $\langle f_m f_n \rangle$, $\langle f_m f_n^\dag \rangle$ and $\langle f_m^\dag f_n \rangle$. From the Jordan-Wigner transformation, we have:
\beq
\begin{cases}
f_n = \prod_{j=1}^{n-1} \sigma_j^z \sigma_n^+, \\
f_n^\dag = \prod_{j=1}^{n-1} \sigma_j^z \sigma_n^-,
\end{cases}
\eeq
where $\sigma_n^+ = \frac{1}{2}\left( \sigma_n^x + i \sigma_n^y \right)$ and $\sigma_n^- = \frac{1}{2}\left( \sigma_n^x - i \sigma_n^y \right)$. Then the four different expectation values become
\beq
\begin{cases}
\langle f_m^\dag f_n^\dag \rangle = \frac{1}{4} \left( C_{m,n}^{xx} - C_{m,n}^{yy} - i C_{m,n}^{xy} - i C_{n,m}^{yx} \right), \\
\langle f_m f_n \rangle = \frac{1}{4} \left( C_{m,n}^{xx} - C_{m,n}^{yy} + i C_{m,n}^{xy} + i C_{n,m}^{yx} \right), \\
\langle f_m^\dag f_n \rangle = \frac{1}{4} \left( C_{m,n}^{xx} + C_{m,n}^{yy} + i C_{m,n}^{xy} - i C_{n,m}^{yx} \right), \\
\langle f_m f_n^\dag \rangle = \frac{1}{4} \left( C_{m,n}^{xx} + C_{m,n}^{yy} - i C_{m,n}^{xy} + i C_{n,m}^{yx} \right),
\end{cases}
\eeq
where:
\beq
C_{m,n}^{\alpha \beta} = \langle \sigma_m^\alpha \prod_{j=m}^{n-1} \sigma_j^z \sigma_n^\beta \rangle,
\eeq
with $\alpha,\beta=x,y$. 

With these general layouts, we show how to explicitly calculate the entanglement entropy of bond fermions by using the $1D$ Kitaev spin chain with two unit cells as an example. The lattice structure of the $1D$ Kitaev spin chain is shown in Fig.~\ref{fig3}(a), where the two unit cells labeled by $0$ and $1$ in purple form the whole lattice. Here the unit cell $1$ (labeled in green) of the left neighbor and the unit cell $0$ (labeled in green) of the right neighbor are included to show the boundaries of the system explicitly. We begin the discussion with the Hamiltonian of the model in momentum space in terms of bond fermions (the derivation of this form can be found in Appendix \ref{appendix a}):
\beq
\tilde{H}_{1D} = \sum_{k} \Psi_{k}^\dag E_k \left(\begin{array}{cc} \cos \theta_k & i\sin \theta_k \\ -i\sin \theta_k & -\cos \theta_k \end{array}\right) \Psi_{k},
\eeq
with $ \Psi_k = \left( b_k, b_{-k}^\dag \right)^T $ and:
\beq
\begin{cases}
E_k = \sqrt{\left( J_z D + J_x \cos k \right)^2 + J_x^2 \sin^2 k}, \\
\cos \theta_k = (J_z D +J_x \cos k)/E_k, \\
\sin \theta_k = J_x \sin k/E_{k}.
\end{cases}
\eeq 
Here $D$ is the local gauge field defined in Eq.~(\ref{interaction_term}).
The Bogoliubov transformation is
\beq
\Psi_{k} = \left(\begin{array}{cc} \cos \frac{\theta_k}{2} e^{-i\varphi_k/2} & \sin \frac{\theta_k}{2} e^{i\varphi_k/2} \\ \sin \frac{\theta_k}{2} e^{-i\varphi_k/2} & -\cos \frac{\theta_k}{2} e^{i\varphi_k/2} \end{array}\right) \left(\begin{array}{c} \tilde{f}_k \\ \tilde{f}_{-k}^\dag \end{array}\right),
\eeq
where $e^{i\varphi_k}=i$ for the model with two unit cells, independent of $k$. This means that $\varphi_k=\pi/2$, independent of $k$. The correlations $\langle f_n f_m^\dag \rangle$, $\langle f_n^\dag f_m \rangle$, $\langle f_n^\dag f_m^\dag \rangle$, and $\langle f_n f_m \rangle$ are then determined from the ground-state expectation values of Bogoliubov fermions
\beq
\langle b_k b_{k'}^\dag \rangle = \delta_{k,k'},~\langle b_k^\dag b_{k'} \rangle=\langle b_k^\dag b_{k'}^\dag \rangle = \langle b_k b_{k'} \rangle = 0.
\eeq

Straightforward calculations show that:
\begin{align}
\langle f_n^\dag f_m^\dag \rangle &= \frac{1}{N_k} \sum_{k} e^{-ik(n-m)} \frac{\sin \theta_k}{2} e^{-i(\varphi_k + \varphi_{-k})/2} \nonumber \\
&= -i\frac{1}{N_k} \sum_{k} e^{-ik(n-m)} \frac{\sin \theta_k}{2}, \nonumber \\
\langle f_n f_m \rangle &= \frac{1}{N_k} \sum_{k} e^{-ik(n-m)} \frac{\sin \theta_k}{2} e^{i(\varphi_k + \varphi_{-k})/2} \nonumber \\
&= i\frac{1}{N_k} \sum_{k} e^{-ik(n-m)} \frac{\sin \theta_k}{2}, \nonumber \\
\langle f_n f_m^\dag \rangle &= \frac{1}{N_k} \sum_k e^{-ik(n-m)} \cos^2 \frac{\theta_k}{2}, \nonumber \\
\langle f_n^\dag f_m \rangle &= \frac{1}{N_k} \sum_k e^{-ik(n-m)} \sin^2 \frac{\theta_k}{2},
\end{align}
where $N_k$ is the number of momentum points in the Brillouin zone. Note that $C_{2m-1,2n-1}$ and $C_{2m,2n}$ are the correlations of the real and imaginary parts of the fermions, respectively. As expected, we find $C_{2m-1,2n-1} = \delta_{m,n}$ and $C_{2m,2n}=\delta_{m,n}$. We also find that
\beq \label{Majorana_correlation}
C_{2m-1,2n} = i \sum_{k} e^{-ik(m-n)} \left( \cos \theta_k + \sin\theta_k \right) = ig_{m-n},
\eeq
and
\beq
C_{2m,2n-1} = - C_{2n-1,2m}.
\eeq
The correlation between the two bond fermions at the unit cells $m$ and $n$ is expressed by the following block matrix:
\beq
\left(\begin{array}{cc} C_{2m-1,2n-1} & C_{2m-1,2n} \\ C_{2m,2n-1} & C_{2m,2n} \end{array}\right) =  \delta_{m,n} I_{2\times2} + i\Xi_{m-n},
\eeq
where:
\beq
\Xi_{m-n} = \left(\begin{array}{cc} 0 & g_{m-n} \\ -g_{-(m-n)} & 0 \end{array}\right).
\eeq
This means that for a subsystem containing $N$ sites, the correlation matrix $\mathcal{C}$ is given by
\beq
\mathcal{C} = I_{2N \times 2N} + i \left(\begin{array}{cccc} \Xi_{0} & \Xi_1 & \cdots & \Xi_{N-1} \\ -\Xi_1^T & \Xi_0 & \cdots & \Xi_{N-2} \\ \vdots & \vdots & \ddots & \vdots \\ -\Xi_{N-1}^T & -\Xi_{N-2}^T & \cdots & \Xi_0 \end{array}\right).
\eeq
For the case with two unit cells, the correlation matrix becomes
\beq
\mathcal{C} = I_{2\times2} + i \Xi_0.
\eeq
Particle-hole symmetry requires $\sin \theta_k = -\sin \theta_{-k}$. This implies that
\beq
g_{0} = \frac{1}{N_k} \sum_{k \in \mathcal{K} \cup -\mathcal{K}} \cos \theta_k.
\eeq 
Using this expression, we obtain the exact curves in black shown in Fig.~\ref{fig3}(c) and (d). To measure the entanglement properties on quantum hardware as a function of momentum, we need to use the symmetry-enforced circuit.

\begin{figure}[tpb]
  \centering
  \includegraphics[width=1\columnwidth]{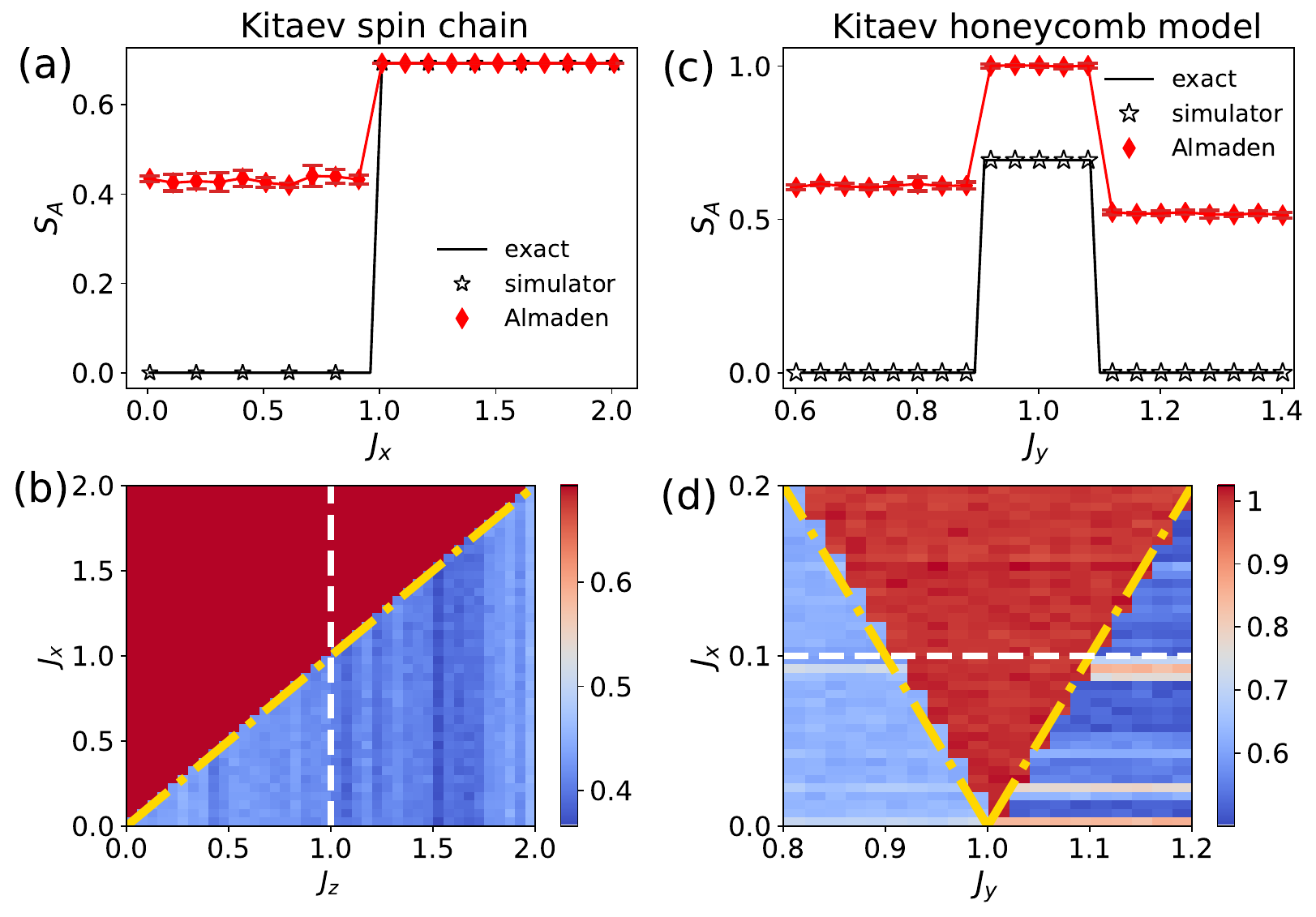}
  \caption{Quantum phase transitions in Kitaev-inspired models determined by entanglement entropy at high symmetry points: (a/c) The entanglement entropy of bond fermions in the subsystem `A' defined in Fig.~2 for the Kitaev spin chain/honeycomb model by measuring the contribution from high-symmetry points. In (a) we set $J_z=1$ and vary $J_x$ to probe different phases, while in (c) we probe different phases by tuning $J_y$ and at the same time setting $J_z=1$ and $J_x=0.1$. In (a) and (c), each data point from simulator was obtained by using $N=8196$ shots, while the results from IBMQ-Almaden are obtained from three independent experiments, with $N=8196$ shots in each experiment. (b/d) Phase diagram of the Kitaev spin chain/honeycomb model determined by the machine IBMQ-Almaden. The gold dashed line denotes the exact phase boundary. In the calculation shown in panel (d), we set $J_z=1$. Each data point in the phase diagrams shown in (b) and (d) was obtained by the IBMQ-Amaden machine with $N=8196$ shots.}
   \label{fig4}
\end{figure}

\subsection{Measuring the entanglement spectrum on quantum hardware by the symmetry-enforce circuit}

We use this symmetry-enforced circuit to obtain the entanglement spectra of the $1D$ Kitaev spin chain containing two unit cells by sweeping across the Brillouin zone and measuring the fermionic correlation functions as described in the subsection \ref{correlation}. The results from the quantum hardware are qualitatively consistent with the exact results. When $J_x<J_z$, the entanglement spectra is always gapped [see Fig.~\ref{fig3}(c)] indicating no edge Majorana modes, while when $J_x>J_z$, the entanglement spectrum is gapless corresponding to the Majorana fermion locating at the virtual boundary between the subsystems [see Fig.~\ref{fig3}(d)]. We highlight that by using this symmetry-enforced circuit, the quantum hardware can even correctly determine the gap closing point at the high-symmetry point $\delta k =\pi$, as shown in Fig.~\ref{fig3}(d).

\section{Determining the quantum phase diagrams on real quantum hardware}
\label{sec:phasediagram}

The high symmetry points with $\delta k=\pi$ in the BZ are the PH symmetric points, and we expect the quantum phase transition to be dominated by the behavior near these points. Hence, we must ensure the simulation includes the effects of these high-symmetry points, in order to efficiently determine the phase diagram on a small cluster. With the help of the symmetry-enforced circuit, this can be done straightforwardly. Using these specific momenta for the systems discussed in Sec. \ref{entanglement_entropy_density_matrix} and Fig.~\ref{fig2}, we obtain the bond fermion entanglement entropy (by measuring correlation functions) as a function of the exchange parameters. 

In Fig.~\ref{fig4}(a), the bond fermion entanglement entropy is shown as a function of exchange strength $J_x$ for the $1D$ Kitaev spin chain with $J_z=1$. We can find that the results from the exact solution, quantum simulators, and the IBMQ-Almaden quantum computer all show a discrete jump, and the entanglement entropy $S_A=\sqrt{2}/2$, when $J_x>J_z$, indicates a Majorana fermion at the virtual boundary of the subsystem. Therefore, the sharp change at $J_x=J_z$ indicates a topological quantum phase transition. Similar situations are found in the bond fermion entanglement entropy for the Kitaev honeycomb model as shown in Fig.~\ref{fig4}(c). Here we set $J_z=1$ and $J_x=0.1$ in the calculations. The sharp jumps in $S_A$ are found to happen at $J_z=J_x+J_y$ and $J_y=J_x+J_z$ and identify two different topological phase transitions of the Kitaev honeycomb model. This efficient identification of the phase boundary relies on the difference of the topology of the single-particle wave functions in the two different phases. When $|J_i|<|J_j+J_k|$ (with $i\neq j\neq k \in \lbrace x,y,z\rbrace$), the single-particle spectrum is gapped, and the ground state of the phase is known as the gapped $Z_2$ spin liquid in the sense that it can be mapped to the toric code model \cite{Kitaev_AP_2006}. On the other hand, when $|J_i|\geq|J_j+J_k|$, the single-particle spectrum is gapless, and the ground state of the phase is a spinon chiral topological superconductor \cite{Jiang_arXiv_2018}. The difference in the single-particle spectrum is revealed by the entanglement entropy of edge modes at the virtual boundaries of the subsystem: if virtual boundaries are introduced by dividing the system into subsystems, then the spinon chiral topological superconducting phase would support edge modes, while the gapped $Z_2$ spin liquid phase would not. Then using this step as the identifying feature, we are able to reconstruct the correct phase diagram of the two models as a function of exchange parameters as shown in Fig.~\ref{fig4}(b) and (d). We have to note that the two gapped phases in the Kitaev honeycomb model have the same properties in their single-particle spectra, but the corresponding quantum states at \emph{the high symmetry points} are quite different and thus expressed by different occupations in the qubit representation. Because the different qubits have different noise levels, the resultant states have quantitatively different entanglement entropy and result in the asymmetry seen in Fig.~\ref{fig4}(d).

\section{Conclusion}
\label{sec:conclusion}

In this work, we illustrate how the topological properties intrinsic to a quantum model being simulated provides topological protection, resulting in robust quantum calculations, even on NISQ hardware. In particular, we demonstrate how to prepare the ground states of these Kitaev-inspired models through low-depth braiding circuits of anyons. By using a particle-hole symmetry-preserving methodology that includes high-symmetry points in the Brillouin zone, the simulation identifies the quantum phase diagram of both the Kitaev spin chain and the original Kitaev honeycomb model for systems larger than are otherwise possible on NISQ hardware (when only local quantities are measured). These results directly show how topological protection results in much more robust calculations on NISQ hardware when preparing and determining different quantum phases. The quantum phase diagram was determined by finding discontinuities in the entanglement entropy, which are reduced in magnitude, but still remained easy to identify in this noisy environment. Hence, the benefits of topological-inspired quantum computation can already be realized on NISQ-era conventional quantum computers and this provides a third methodology for topological quantum computation that explores a different realm than quantum error correction or topological quantum computing with topological qubits.

\section*{Acknowledgements}

This work was supported by the Department of Energy, Office of Basic Energy Sciences, Division of Materials Sciences and Engineering under Grant No. DE-SC0019469. J.K.F. was also supported by the McDevitt bequest at Georgetown.
We acknowledge use of the IBM Q for this work. The views expressed are those of the authors and do not reflect the official policy or position of IBM or the IBM Q team. Access to the IBM Q Network was obtained through the IBM Q Hub at NC State. We acknowledge the use of the QISKIT software package~\cite{qiskit2019} for performing the quantum simulations. The authors also thank Sonika Johri for insightful comments and Akhil Francis for some technical expertise. The codes and data are available at \href{https://datadryad.org/stash/share/TeNf4pRWHHIGmxlb0HZyuNBRWjizViBVWqMuTrYRcFA}{https://doi.org/10.5061/dryad.0zpc866vk}.

\appendix

\section{The exact solution of the $1D$ Kitaev spin chain} \label{appendix a}

The Hamiltonian of the $1D$ Kitaev spin chain is given by
\beq
H_{1D} = 4J_z \sum_{i=1}^{N/2} \sigma_{2i-1}^z \sigma_{2i}^z + 4J_x \sum_{i=1}^{N/2} \sigma_{2i}^x \sigma_{2i+1}^x. 
\eeq
Here we assume that the odd sites belong to the $\bullet$ sublattice while the even sites belong to the $\circ$ sublattice (the configuration as shown by the inset in Fig.~\ref{fig2}(a)). The factor of $4$ is added for convenience, as we will see below. To solve this model, we first map the spin model into a fermionic model by the Jordan-Wigner transformation:
\beq
\begin{cases}
\sigma_m^+ \to \prod_{j<m} \sigma_j^z c_m, \\
\sigma_m^- \to \prod_{j<m} \sigma_j^z c_m^\dag, \\
\sigma_m^z \to \frac{1}{2}\left(1 - 2c_m^\dag c_m\right),
\end{cases}
\eeq
where we use the standard mapping $|0\rangle=|\uparrow\rangle$ and $|1\rangle=|\downarrow\rangle$. After fermionization, the Hamiltonian becomes
\begin{align}
\tilde{H}_{1D} &= J_z \sum_i^{N/2} \left( 2 n_{2i-1} - 1 \right) \left( 2n_{2i} - 1 \right) \nonumber \nonumber \\
& + J_x \sum_{i=1}^{N/2} \left( c_{2i} - c_{2i}^\dag \right) \left( c_{2i+1} + c_{2i+1}^\dag \right),
\end{align}
which is identical to the fermionic expression of the Kitaev honeycomb model in the limit $J_y=0$ \cite{Kitaev_AP_2006}. To solve this, we must transform the four-fermion term. We 
introduce the Majorana fermions on each sublattice $\eta_{\circ/\bullet}$ and $\gamma_{\circ/\bullet}$ \cite{Feng_PRL_2007,Schmoll_PRB_2017}:
\beq
\begin{cases}
c_{\circ} = \left( \eta_{\circ} + i\gamma_{\circ} \right)/2, \\
c_{\bullet} = \left( \gamma_{\bullet} + i\eta_{\bullet} \right)/2.
\end{cases}
\eeq
The Hamiltonian is re-expressed as
\beq
\tilde{H}_{1D} = -iJ_x \sum_{i} \gamma_{i,\circ} \gamma_{i+1,\bullet} + iJ_z \sum_{i} D_i \gamma_{i,\bullet} \gamma_{i,\circ},
\eeq
where $D_i = i\eta_{i,\bullet} \eta_{i,\circ}$ \cite{Kitaev_AP_2006,Feng_PRL_2007} is a local gauge field. Here $i$ labels unit cells, which have two sublattices $\bullet$ and $\circ$. Since the Majorana fermions do not have well-defined occupation numbers, we construct quantum circuits after mapping the Majorana fermions back to conventional fermions. We introduce \emph{bond fermions} for the two sites connected by the $z$-bonds as follows:
\beq
f_{i} = \frac{1}{2} \left( \gamma_{i,\bullet} + i\gamma_{i,\circ}\right).
\eeq
Finally, we re-express the Hamiltonian again and find
\begin{align} \label{Eq:fermion_A}
\tilde{H}_{1D} &= J_x \sum_{i} \left( f_i^\dag -f_i \right) \left( f_{i+1}^\dag +f_{i+1} \right) \nonumber \\
&+ J_z \sum_i D_i (2f_i^\dag f_i - 1). 
\end{align}
Because of the translational symmetry, Lieb's theorem indicates that the fluxes $D_i$ must be uniform \cite{Kitaev_AP_2006}. Since $D_i$ commutes with the Hamiltonian $\tilde{H}_{1D}$, it is a constant of motion. Note that $\eta_{\circ}^2 = \eta_{\bullet}^2=1$, so $D_i=\pm1$.  

To diagonalize the Hamiltonian, we start with a Fourier transformation
\beq
f_{n} = \frac{1}{\sqrt{N}} \sum_{k} e^{i k n} \tilde{f}_{k},
\eeq
where the Hamiltonian becomes
\beq
\tilde{H}_{1D} = \sum_{k} \Psi_{k}^\dag \left(\begin{array}{cc} J_z D + J_x \cos k & iJ_x \sin k \\ -i J_x \sin k & -J_z D - J_x \cos k \end{array}\right) \Psi_{k},
\eeq
after dropping some irrelevant constants. Here, we introduce the ``spinor"
\beq
\Psi_{k} = \left(\begin{array}{c} \tilde{f}_k \\ \tilde{f}_{-k}^\dag \end{array} \right).
\eeq
This Hamiltonian in momentum space can be easily diagonalized through a Bogoliubov transformation. Doing so gives us the final diagonal form
\beq
\tilde{H}_{1D} = \sum_{k} E_k \left( b_{k}^\dag b_{k} - b_{-k} b_{-k}^\dag \right),
\eeq
where $E_k = \sqrt{(J_z D + J_x \cos k)^2 + J_x^2 \sin^2 k}$, and $b_{k}$ ($b_{k}^\dag$) denotes the annihilation (creation) operator for Bogoliubov fermions with momentum $k$.

\section{The Bogoliubov and Fourier circuit for the Kitaev honeycomb model} \label{circuit_Bogoliubov_Fourier}

\subsection{The implementation of Bogoliubov transformation}

The first step is to reverse the Bogoliubov transformation to transform the representation formed by bond fermions in momentum space. In general, the Bogoliubov transformation is a two-qubit operation \cite{Verstraete_PRA_2009,Lierta_Qu_2018}. By assuming that the order of qubits is from bottom to top, the quantum circuit to realize this two-qubit Bogoliubov transformation is drawn in the right middle panel in Fig.~\ref{fig1}(c). Performing the pairing between opposite momentum points, the Bogoliubov transformation for the Kitaev honeycomb model is achieved by the two two-qubit Bogoliubov transformations (denoted by orange boxes in Fig.~\ref{fig1}(b)) operating on neighboring qubits.

\subsection{The implementation of Fourier transformation}

We next convert from the bond fermion state in momentum space to real space by implementing two Fourier transformations sequentially along the $y$ and $x$ directions. Based on the fast Fourier transform method, the Fourier transformation for one of the directions containing $N=2^m$ qubits ($m$ an integer) can be realized with a log-depth circuit and using at most two-qubit quantum gates \cite{Lierta_Qu_2018,Ferris_PRL_2014}. For the Kitaev honeycomb model on the lattice containing a $2 \times 2=4$ unit cells, the Fourier transformation along the $y$-direction consists of two two-qubit Fourier transformations for $k_x=\pi/2$ and $k_x=-\pi/2$, and the transformation along the $x$-direction also contains two two-qubit transformations for $k_y=\pi/2$ and $k_y=-\pi/2$ respectively. In particular, the two-qubit Fourier transformation consists of a phase delay factor implemented by the single-qubit gate $U_1(\phi)$ shown in the right upper panel in Fig.~\ref{fig1}(c) and a two-qubit `beam-splitter' operation implemented by the rest of the right upper panel in Fig.~\ref{fig1}(c) except for $U_1(\phi)$. Beginning from the states obtained after the Bogoliubov transformation, we need to reorder the qubits according to the momentum points as $(-\pi/2,-\pi/2)$, $(-\pi/2,\pi/2)$, $(\pi/2,-\pi/2)$, and $(\pi/2,\pi/2)$, which is achieved by the two swap operations after the Bogoliubov boxes in Fig.~\ref{fig1}(b), so that the Fourier transformation in the $y$-direction can be implemented for the two sets of neighboring qubits, respectively, shown by the two green boxes following the two swap operations in Fig.~\ref{fig1}(b). To facilitate the Fourier transformation in the $x$-direction, we need to switch the order of the two qubits for $(-\pi/2,\pi/2)$ and $(\pi/2,-\pi/2)$, which is achieved by the swap operation following the two green boxes in Fig.~\ref{fig1}(b). In this way, the Fourier transformation along the $x$-direction can be implemented by the two green boxes following the swap box as shown in Fig.~\ref{fig1}(b).

\section{The braiding circuit of the $1D$ Kitaev spin chain} \label{appendix b}

As we have discussed in the main text, the details of the braiding circuit are determined by the constraints set by local coupling terms of the Hamiltonian. The constraint set by the $z$-bond is general for the Kitaev-inspired models, so here we will focus on the constraints from other coupling terms in the $1D$ Kitaev spin chain.

To make the discussion explicit, we focus on the $x$-bond connecting sites $2$ and $3$. The original Hamiltonian connecting across the $x$-bond is
\beq
H_x = J_x \left( -c_2 + c_2^\dag \right) \left( c_3 + c_3^\dag \right),
\eeq
and it is re-expressed in terms of bond and gauge fermions as
\beq
\tilde{H}_x = J_x \left( f_1^\dag - f_1 \right) \left( f_2^\dag + f_2 \right),
\eeq
where the braiding operations on $c_1$ and $c_2$ give the bond fermion $f_1$ and the gauge fermion $g_1$, and similarly the braiding operations on $c_3$ and $c_4$ give the bond fermion $f_2$ and the gauge fermion $g_2$. Hence, the states considered here contain four qubits. Start with the $n_{g_1}=n_{g_2}=1$ case. The relevant states for $\tilde{H}_x$ are $|n_{f_1}n_{f_2}n_{g_1}n_{g_2}\rangle = |n_{f_1}n_{f_2}11\rangle_f$. They are transformed under $\tilde{H}_x$ as
\begin{align} \label{x_coupling_bond_gauge}
\tilde{H}_x |0011\rangle_f = J_x |1111\rangle_f,~\tilde{H}_x |1111\rangle_f = J_x |0011\rangle_f, \nonumber \\
\tilde{H}_x |0111\rangle_f = J_x |1011\rangle_f,~\tilde{H}_x |1011\rangle_f = J_x |0111\rangle_f.
\end{align}
From the general form of braiding provided by Eq.~(\ref{braiding}), the states $|n_{f_1}n_{f_2}11\rangle_f$ are transformed to $|n_1 n_2 n_3 n_4\rangle_c$ (in the representation of fermions denoted by $c$ operators) as follows:
\begin{align}
\mathcal{U}_B |0011\rangle_f &= (\alpha_{01}^{01})^2 |0101\rangle_c + \alpha_{01}^{10} \alpha_{01}^{01} |1001\rangle_c \nonumber \\
&+ \alpha_{01}^{01}\alpha_{01}^{10}|0110\rangle_c + (\alpha_{01}^{10})^2 |1010\rangle_c, \nonumber \\
\mathcal{U}_B |1111\rangle_f &= -(\alpha_{11}^{00})^2 |0000\rangle_c - \alpha_{11}^{00}\alpha_{11}^{11} |0011\rangle_c \nonumber \\
&- \alpha_{11}^{11}\alpha_{11}^{00} |1100\rangle_c - (\alpha_{11}^{00})^2 |1111\rangle_c, \nonumber \\
\mathcal{U}_B |0111\rangle_f &= -\alpha_{01}^{01} \alpha_{11}^{00} |0100\rangle_c -\alpha_{01}^{01} \alpha_{11}^{11} |0111\rangle_c \nonumber \\
&-\alpha_{01}^{10} \alpha_{11}^{00} |1000\rangle_c - \alpha_{01}^{10} \alpha_{11}^{11} |1011\rangle_c, \nonumber \\
\mathcal{U}_B |1011\rangle_f &= \alpha_{11}^{00} \alpha_{01}^{01} |0001\rangle_c + \alpha_{11}^{00} \alpha_{01}^{10} |0010\rangle_c \nonumber \\
&+\alpha_{11}^{11} \alpha_{01}^{01} |1101\rangle_c + \alpha_{11}^{11} \alpha_{01}^{10} |1110\rangle_c.
\end{align}
The action of $H_x$ on these states gives
\begin{align} \label{x_coupling_conventional}
H_x |0101\rangle_c &= J_x | 0011 \rangle_c,~H_x |1001\rangle_c = J_x |1111\rangle_c, \nonumber \\
H_x |0110\rangle_c &= J_x |0000\rangle_c,~H_x|1010\rangle_c = J_x|1100\rangle_c, \nonumber \\
H_x |0100\rangle_c &= J_x |0010\rangle_c,~H_x |0111\rangle_c = J_x |0001\rangle_c, \nonumber \\
H_x |1000\rangle_c &= J_x |1110\rangle_c,~H_x |1011\rangle_c = J_x |1101\rangle_c.
\end{align}
The consistency between $\langle n_{f_1} n_{f_2} 1 1| \tilde{H}_x | n_{f_1} n_{f_2} 1 1 \rangle_f$ and $\langle n_{f_1} n_{f_2} 1 1| \mathcal{U}_B^\dag H_x \mathcal{U}_B | n_{f_1} n_{f_2} 1 1 \rangle_f$ implies
\begin{align} \label{coef_relation}
&(\alpha_{01}^{01})^2 = (\alpha_{01}^{10})^2 = -\alpha_{11}^{00}\alpha_{11}^{11},~(\alpha_{11}^{00})^2 = (\alpha_{11}^{11})^2 = -\alpha_{01}^{01}\alpha_{01}^{10}, \nonumber \\
&\alpha_{01}^{01}=-\alpha_{01}^{10},~\alpha_{11}^{00}=-\alpha_{11}^{11}.
\end{align}
This yields $\alpha_{01}^{01}=\pm \frac{i}{\sqrt{2}}$, $\alpha_{01}^{10}=\mp \frac{i}{\sqrt{2}}$, $\alpha_{11}^{00}=\pm \frac{i}{\sqrt{2}}$, and $\alpha_{11}^{11}=\mp \frac{i}{\sqrt{2}}$. Since $\alpha_{01}^{01}$ and $\alpha_{11}^{11}$ are imaginary, there are two intra-fermion braidings acting on the same fermion to adjust the global phase by $\pm i$. The inter-fermion braiding introduces a $\pi/2$ phase difference. To increase the phase difference to  $\pi$, one of the intra-fermion braidings is performed after the inter-fermion braiding. We still need to determine on which fermion the intra-fermion braiding acts. By using Eq.~(\ref{coef_relation}), the transformation between the two representation can be written as:
\beq \label{transformation}
\begin{cases}
\mathcal{U}_B |01\rangle_f = -\alpha_{11}^{11} |01\rangle_c + \alpha_{11}^{11} |10\rangle_c, \\
\mathcal{U}_B |11\rangle_f = \alpha_{11}^{11} |11\rangle_c - \alpha_{11}^{11} |00\rangle_c.
\end{cases}
\eeq
But, the phase difference introduced by the inter-fermion braiding operations are \emph{opposite} for $|01\rangle$ and $|11\rangle$. To keep the form of Eq.~(\ref{transformation}) up to a global minus sign, the only choice is that the intra-fermion braiding operation is applied to the first qubit, the occupation of which is different for the two states.
\begin{figure}[htpb]
  \centering
  \includegraphics[width=0.3\columnwidth]{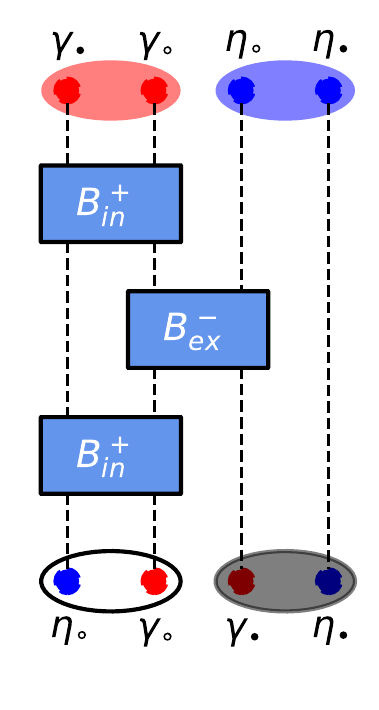}
  \caption{The braiding circuit used for the $1D$ Kitaev spin chain.}
   \label{figS1}
\end{figure}

From the analysis provided above, the braiding circuit must satisfy the following:

1. the circuit is constructed by one inter-fermion braiding and two intra-fermion braidings;

2. one of the intra-fermion braidings is performed after the inter-fermion braiding;

3. the two intra-fermion braiding operations act on the first qubit.

A circuit fulfilling these requirements is shown in Fig.~\ref{figS1}. One can apply a similar analysis to the case with $n_g=0$. One finds the same requirements in that case.

\section{The $1D$ BCS-Hubbard model and the circuit construction} \label{appendix c}

\begin{figure}[htpb]
  \centering
  \includegraphics[width=0.7\columnwidth]{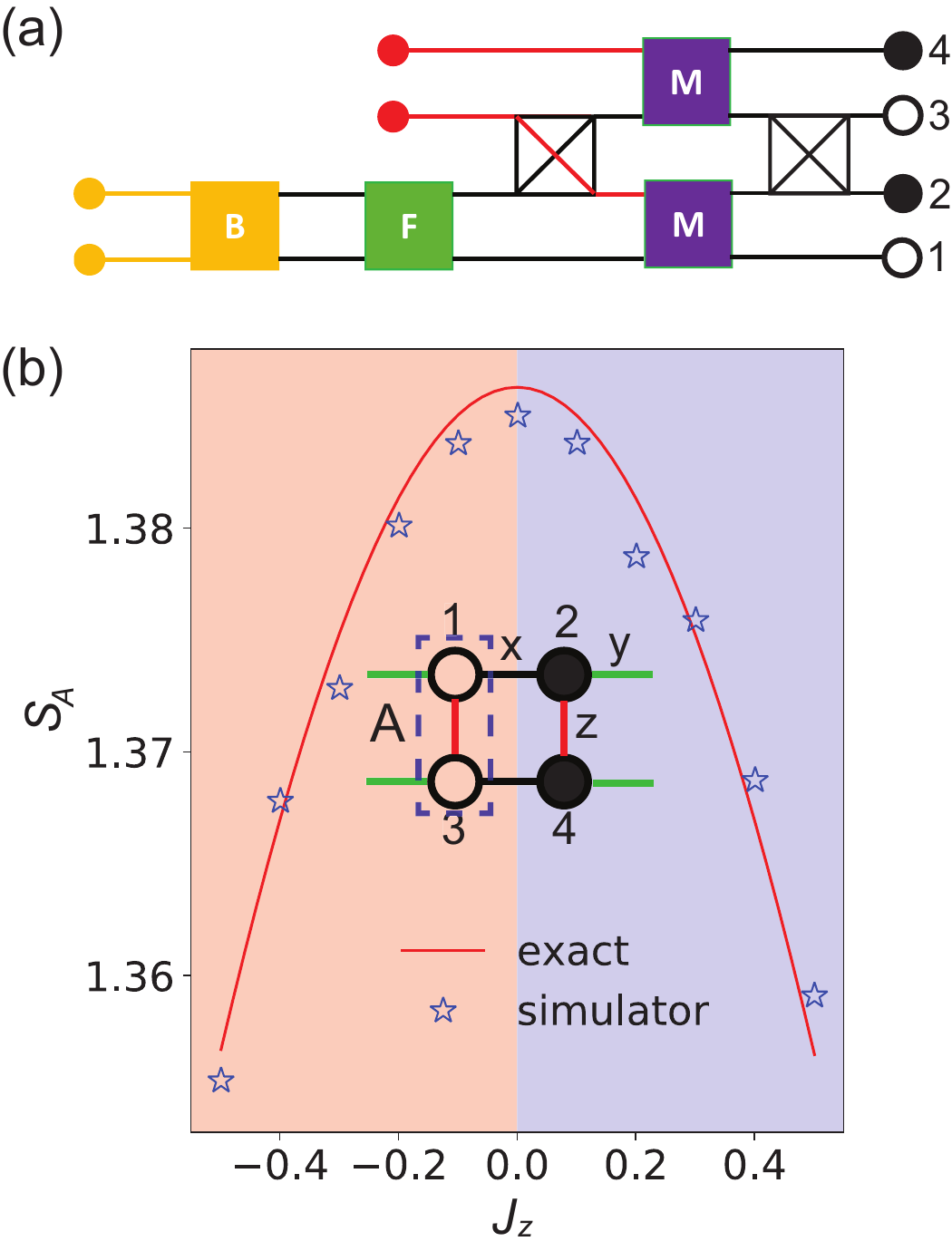}
  \caption{(a) the quantum circuit to prepare the ground states of $1D$ BCS-Hubbard model; (b) the entanglement entropy of the ground state of the $1D$ BCS-Hubbard model containing a lattice of two unit cells (four sites overall) is shown in the inset. The part in the dashed box is the subsystem $A$, and the entanglement entropy is calculated for this subsystem, and $\text{x}$, $\text{y}$, and $\text{z}$ denote the $x$-bonds in black, $y$-bonds in green and $z$-bonds in red. We use $N=8196$ in the calculation of entanglement entropy with the quantum simulator.}
   \label{figS2}
\end{figure}

The Hamiltonian of the $1D$ BCS-Hubbard model can be written as:
\begin{align}
H &= J_x \sum_{l=1,2} \sum_{i=1}^{N/2} \left(\sigma_{2i-1,l}^x \sigma_{2i,l}^x + \sigma_{2i,l}^y \sigma_{2i+1,l}^y\right) \nonumber \\
&+ J_z \sum_{i} \sigma_{i,1}^z \sigma_{i,2}^z,
\end{align}
which is constructed on a ladder lattice as shown in the inset of Fig.~\ref{figS2}(b) with the subscript $l$ denoting the two legs of the ladder. The JW string is defined so that the odd sites belongs to the $\circ$ sublattice and the even ones are the $\bullet$ sublattice. Obviously, this model is a spin $1/2$ model fulfilling the requirements stated in \cite{Miao_arXiv_2018} and thus another example of Kitaev-inspired model.

By the Jordan-Wigner transformation, we have that
\beq
\begin{cases}
\sigma_m^+ \to \prod_{j<m} \sigma_j^z c_m, \\
\sigma_m^- \to i\prod_{j<m} \sigma_j^z c_m^\dag, \\
\sigma_m^z \to 1 - 2c_m^\dag c_m,
\end{cases}
\eeq
where we have identified $|0\rangle=|\uparrow\rangle$ and $|1\rangle=|\downarrow\rangle$. Hence, we find: $c_{i}^\dag \sigma_i^z = c_{i}^\dag$ and $c_{i} \sigma_i^z = -c_{i}$:
\beq
\sigma_{2i-1,l}^x \sigma_{2i,l}^x = \left( -c_{2i-1,l} + c_{2i-1,l}^\dag \right) \left( c_{2i,l} + c_{2i,l}^\dag \right).
\eeq
Similarly:
\beq
\sigma_{2i,l}^y \sigma_{2i+1,l}^y = -\left( c_{2i,l} + c_{2i,l}^\dag \right) \left( -c_{2i+1,l} + c_{2i+1,l}^\dag \right).
\eeq
Inserting the Jordan-Wigner transformation for $\sigma^z$, we find the fermionic Hamiltonian:
\begin{align} \label{Hamiltonian_c}
\tilde{H} &= J_x \sum_{i\in\circ,l} \left( c_{i,l}^\dag c_{i+1,l} + c_{i,l}^\dag c_{i+1,l}^\dag  + h.c.\right) \nonumber \\
&+J_x \sum_{i\in\circ,l} \left( c_{i,l}^\dag c_{i-1,l} + c_{i,l}^\dag c_{i-1,l}^\dag + h.c.\right) \nonumber \\
&+ 4J_z \sum_{i} \left( n_{i,1}-1/2 \right) \left( n_{i,2}-1/2 \right),
\end{align}
where we used the fact that all the odd sites are $\circ$ sites, and all the even sites are $\bullet$ sites. To simplify the four-fermion interaction term, we assume that the fermions at the $\bullet$ sites are decomposed into Majorana fermions by following:
\beq
c_{\bullet,l} = (\gamma_{\bullet,l} - i \eta_{\bullet,l})/2,
\eeq
while the fermions at the $\circ$ sites are decomposed as:
\beq
c_{\circ,l} = (\eta_{\circ,l} + i \gamma_{\circ,l})/2.
\eeq
Then the Hamiltonian can be expressed by Majorana fermions:
\begin{align}
\tilde{H} = &i J_x \sum_{i\in \circ,l} \left(\gamma_{i,l} \gamma_{i+1,l} + \gamma_{i,l} \gamma_{i-1,l}\right) \nonumber \\
&- i J_z \sum_{i} (i\eta_{i,1} \eta_{i,2}) \gamma_{i,1} \gamma_{i,2}.
\end{align}
The quantities $D_{i} = i\eta_{i,1} \eta_{i,2}$ commute with the Hamiltonian and have the eigenvalues $D_{i}=\pm1$. Then we need to express the model in terms of bond fermions, and we introduce the following transformations:
\beq
\begin{cases}
f_{\circ} = (\gamma_{\circ,1} + i \gamma_{\circ,2})/2,~g_{\circ} = (\eta_{\circ,2} + i \eta_{\circ,1})/2, \\
f_{\bullet} = (\gamma_{\bullet,1} - i \gamma_{\bullet,2})/2,~g_{\bullet} = (\eta_{\bullet,2} - i \eta_{\bullet,1})/2.
\end{cases}
\eeq
Then the Hamiltonian (in terms of fermions defined in above) can be written as:
\begin{align} \label{Hamiltonian_f}
\tilde{H} &= i 2J_x \sum_{i \in \circ} \left( f_{i}^\dag f_{i+1}^\dag + f_{i}^\dag f_{i-1}^\dag - h.c. \right) \nonumber \\
& + \sum_{i} J_z \tilde{D}_{i} \left( 2 f_{i}^\dag f_{i} - 1 \right),
\end{align}
where $\tilde{D}_{i}=-D_{i}$ for $i\in\circ$, and $\tilde{D}_{i}=D_{i}$ for $i\in\bullet$. According to Lieb's theorem and direct numerical verifications \cite{Kitaev_AP_2006, Chen_PRL_2018}, $\tilde{D}_{i} = D$ should be uniform to minimize the energy of the system. To further diagonalize the Hamiltonian, we Perform Fourier transformation for the two sublattice as follows:
\beq
\begin{cases}
f_{n,\circ} = \frac{1}{\sqrt{N}} \sum_{k} e^{-i k n_{\circ}} f_{k,\circ}, \\
f_{n,\bullet} = \frac{1}{\sqrt{N}} \sum_{k} e^{-i k n_{\bullet}} f_{k,\bullet},
\end{cases}
\eeq
where $n$ labels the unit cell, $n_{\circ}$ and $n_{\bullet}$ denote the position of the $\circ$ and $\bullet$ in the unit cell labeled by $n$. We set the lattice constant $a=1$, and the distance between $\circ$ and $\bullet$ sites in the unit cell is $1/2$. The Hamiltonian in momentum space can then be written as
\beq
\tilde{H} = \sum_{k} \Psi_{k}^\dag \left(\begin{array}{cc} 2J_z D & -iJ_x \cos\frac{k}{2} \\ iJ_x \cos \frac{k}{2} & -2J_z D \end{array}\right) \Psi_{k},
\eeq
where
\beq
\Psi_{k}=\left(\begin{array}{c} f_{k,\circ} \\ f_{-k,\bullet}^\dag \end{array}\right).
\eeq
Then the single-particle Hamiltonian can be further simplified to
\beq
\tilde{H} = \sum_{k} \left( E_{k,+} b_{k}^\dag b_{k} + E_{k,-} b_{-k} b_{-k}^\dag\right),
\eeq
where $b_k$ denotes the Bogoliubov fermion, and
\beq
E_{k,\pm} = \pm \sqrt{ J_x^2 \cos^2\frac{k}{2} + 4J_z^2 }.
\eeq

Based on the exact solution of the model described above, the quantum circuit to prepare the ground state of the model can be constructed by following the procedures given in Sec.~\ref{kitaev_circuit}, and the resulting quantum circuit is shown in Fig.~\ref{figS2}~(a). The correctness of the quantum circuit is verified by calculating the entanglement entropy in a quantum simulator for the subsystem $A$ defined in the inset of Fig.~\ref{figS2}. The simulator result only differs from the exact result by statistical errors.

By comparing the Hamiltonian of this model, written in terms of bond fermions given by Eq.~(\ref{Hamiltonian_f}), with the Hamiltonian of the $1D$ Kitaev spin chain, written in terms of bond fermions given by Eq.~(\ref{Eq:fermion_A}), the only difference is that there are no fermionic hopping terms in the $1D$ BCS-Hubbard model. Therefore, the entanglement entropy of bond fermions in the $1D$ BCS-Hubbard model (containing two unit cells) can be measured by the same circuit as used for the $1D$ Kitaev spin chain (containing two unit cells) as shown in Fig.~\ref{fig3} in the main text. Indeed, both models have  particle-hole symmetry and are classified as $1D$ $D$-class superconductors according to the classification of non-interacting topological insulators \cite{Schnyder_PRB_2008}. A $\mathbb{Z}_2$ invariant \cite{Schnyder_PRB_2008} can be defined at the high-symmetry points of both models:
\beq
\nu = \mathrm{sgn}\left[ \mathrm{Pf}\left( \tilde{H}(k=0) \right) \mathrm{Pf}\left( \tilde{H}(k=\pi) \right) \right].
\eeq
This topological invariant describes the twist of the energy bands due to the gap closing at the high-symmetry points. For the $1D$ BCS-Hubbard model, the sub-lattice symmetry make the $k=0$ points  always gapped, so $\mathrm{Pf} [\tilde{H}(k=0)]$ does not change sign. This suggests that the quantum phase transition in the $1D$ BCS-Hubbard model depends on the sign of $\mathrm{Pf}[\tilde{H}_{B}(k=\pi)]$, which is equal to the sign of $J_z$. Therefore, the quantum phase transition in the $1D$ BCS-Hubbard model happens when $J_z$ changes sign, which is consistent with the numerical results shown in \cite{Chen_PRL_2018}. A straightforward check is that when $J_z$ changes sign, the filling at $k=\pi$ changes as we expect it would.

\section{The correctness of the ground state preparation} \label{appendix d}

\begin{figure}[!h]
  \centering
  \includegraphics[width=1\columnwidth]{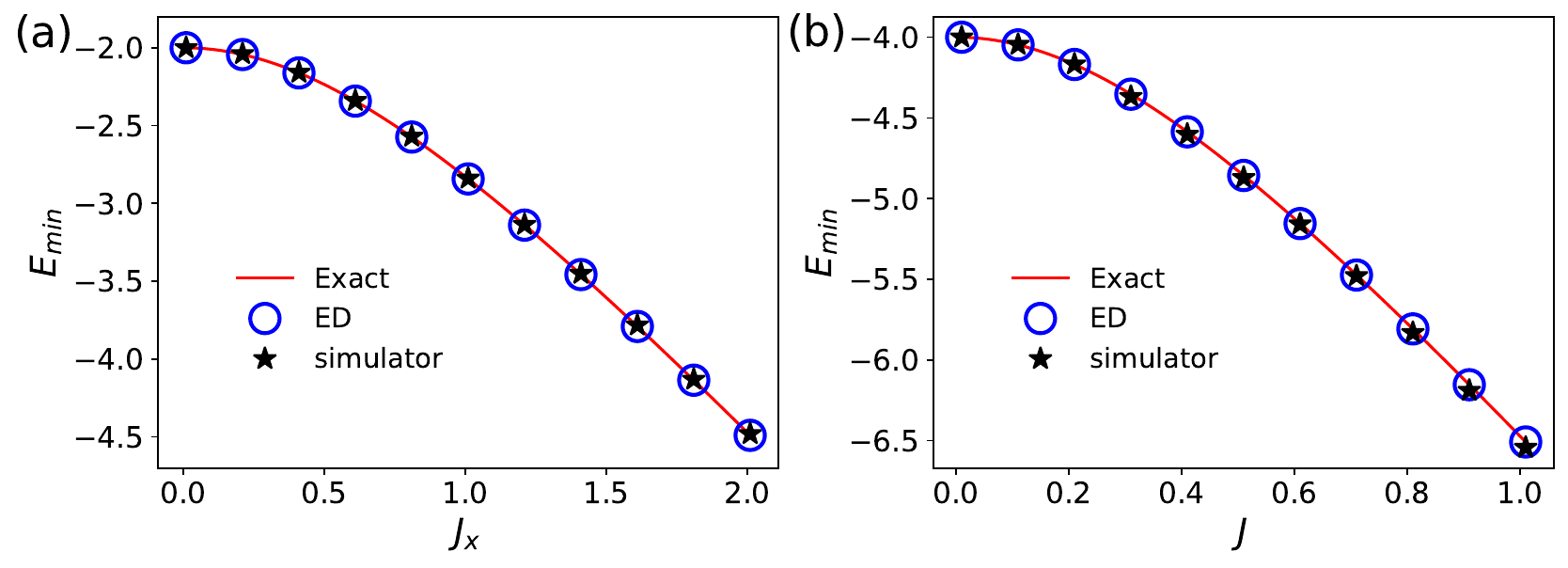}
  \caption{Verification of the ground-state preparation: (a) the $1D$ Kitaev spin chain and (b) the Kitaev honeycomb model. For the $1D$ Kitaev spin chain, we set $J_z=1$ to determine the lowest energy as a function of $J_x$. For the Kitaev honeycomb model, we set $J_z=1$ and $J_x=J_y=J$ to determine the lowest energy as a function of $J$. Each data point from quantum simulators is obtained by performing $N=8196$ experiments.}
   \label{figS3}
\end{figure}

In this section, we prove that the quantum circuit proposed in the main text (Fig.~\ref{fig1}) correctly prepares the ground state of Kitaev-inspired models. The system sizes of the models considered here are the same as those shown in the main text. The lowest energy of the finite-size cluster is determined by exact diagonalization, and then  compared to the results obtained by exact solution and by quantum circuit simulation. Results for the $1D$ Kitaev spin chain and the Kitaev honeycomb model are summarized in Fig.~\ref{figS3}. It turns out that the lowest energies obtained from exact diagonalization are identical to those obtained from the exact solution. Results from the quantum simulators have small deviations due to finite-number sampling.

\section{Finite size effect on entanglement entropy} \label{appendix e}

\begin{figure}[!h]
  \centering
  \includegraphics[width=0.8\columnwidth]{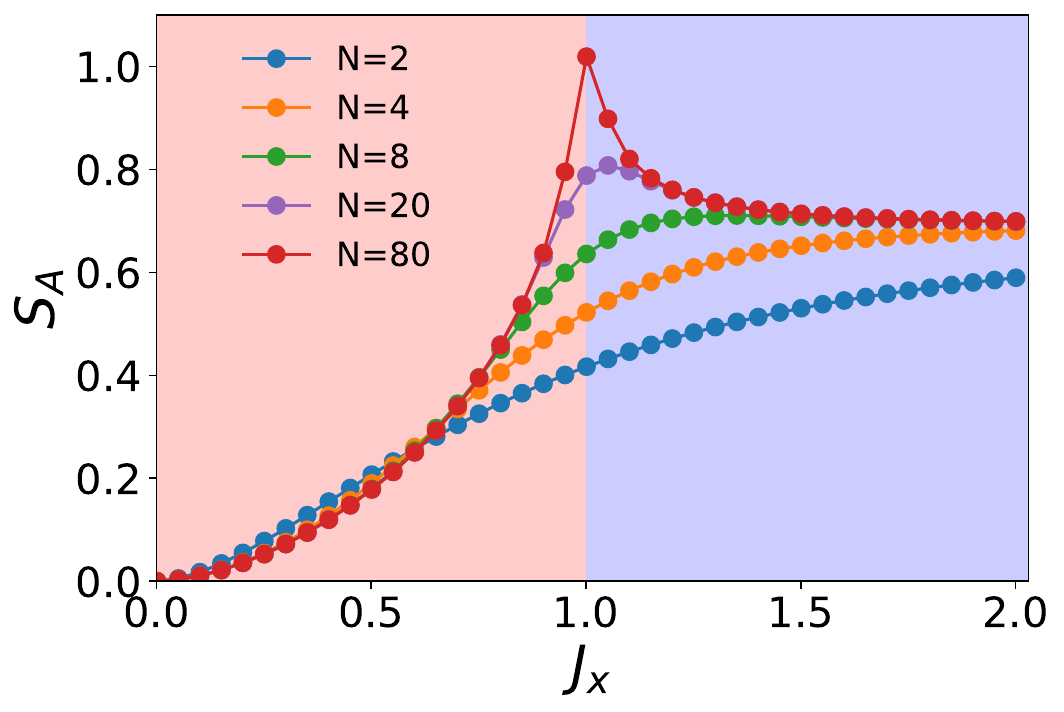}
  \caption{The finite-size effect on the entanglement entropy of the $1D$ Kitaev spin chain: the signature of the quantum phase transition appears when the size of the lattice formed by bond fermions $N$ is significantly larger than $10$. The position of quantum phase transition point can be found accurately when $N\geq80$. We define the subsystem $A$ to be the equal partition of the lattice, and $J_z$ is set to be $1$.}
   \label{figS4}
\end{figure}

\begin{figure}[!h]
  \centering
  \includegraphics[width=0.8\columnwidth]{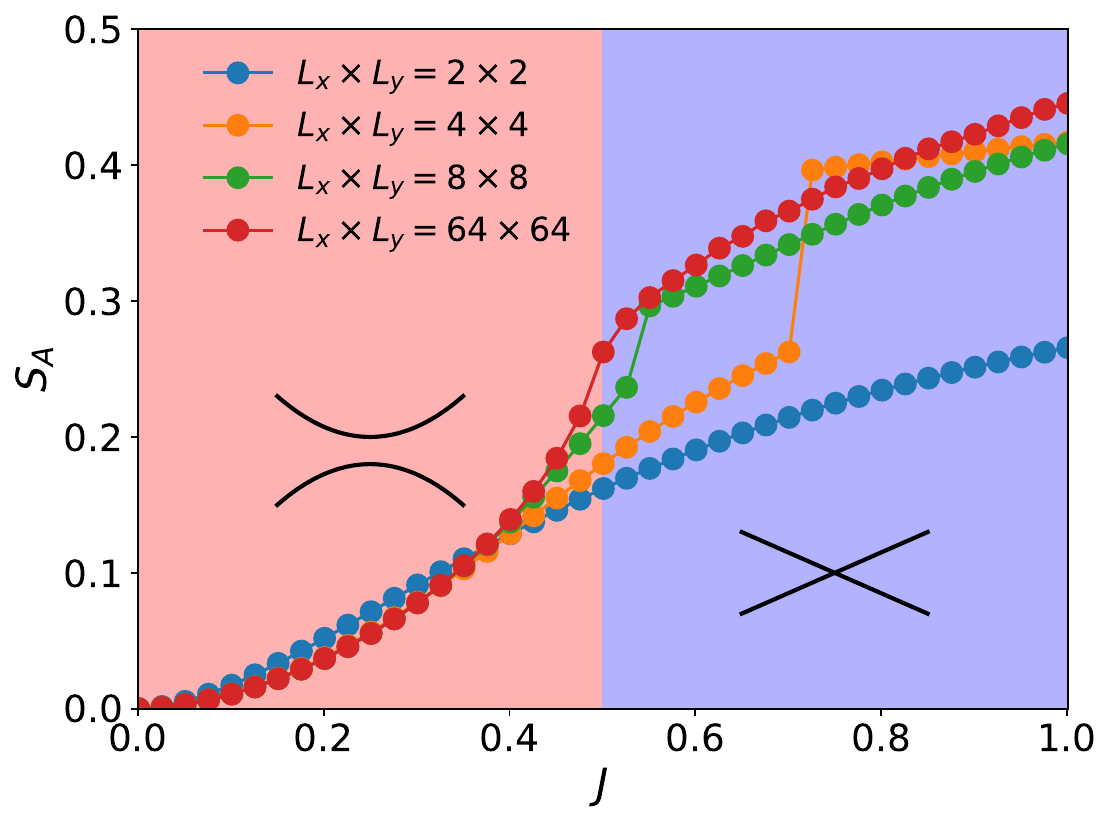}
  \caption{The finite-size effects on the entanglement entropy of the Kitaev honeycomb model. The position of the quantum phase transition can be determined only when $L_x\times L_y\geq64 \times 64$. Subsystem $A$ is defined to be the equal partition of the lattice, and $J_z=1$ and $J_x=J_y=J$.}
   \label{figS5}
\end{figure}

The advantage of calculating entanglement entropy with correlation functions is that it allows us to study the finite-size effects on the entanglement entropy. In Fig.~\ref{figS4}, we show how the entanglement entropy of the bond fermions changes with increasing lattice size. We note that when the bond fermion lattice size $N$ is smaller than $10$, the entanglement entropy increases with the coupling strength $J_x$ monotonically, and no obvious signature appears when $J_x$ passes through the quantum phase transition point. If we further increase the lattice size (for example the $N=20$ case), the non-monotonic behavior of the entanglement entropy with increasing coupling strength starts to appear. However, an accurate determination of the quantum phase transition requires the bond fermion lattice size $N$ to be larger than about $80$.

\begin{figure}[!ht]
  \centering
  \includegraphics[width=0.9\columnwidth]{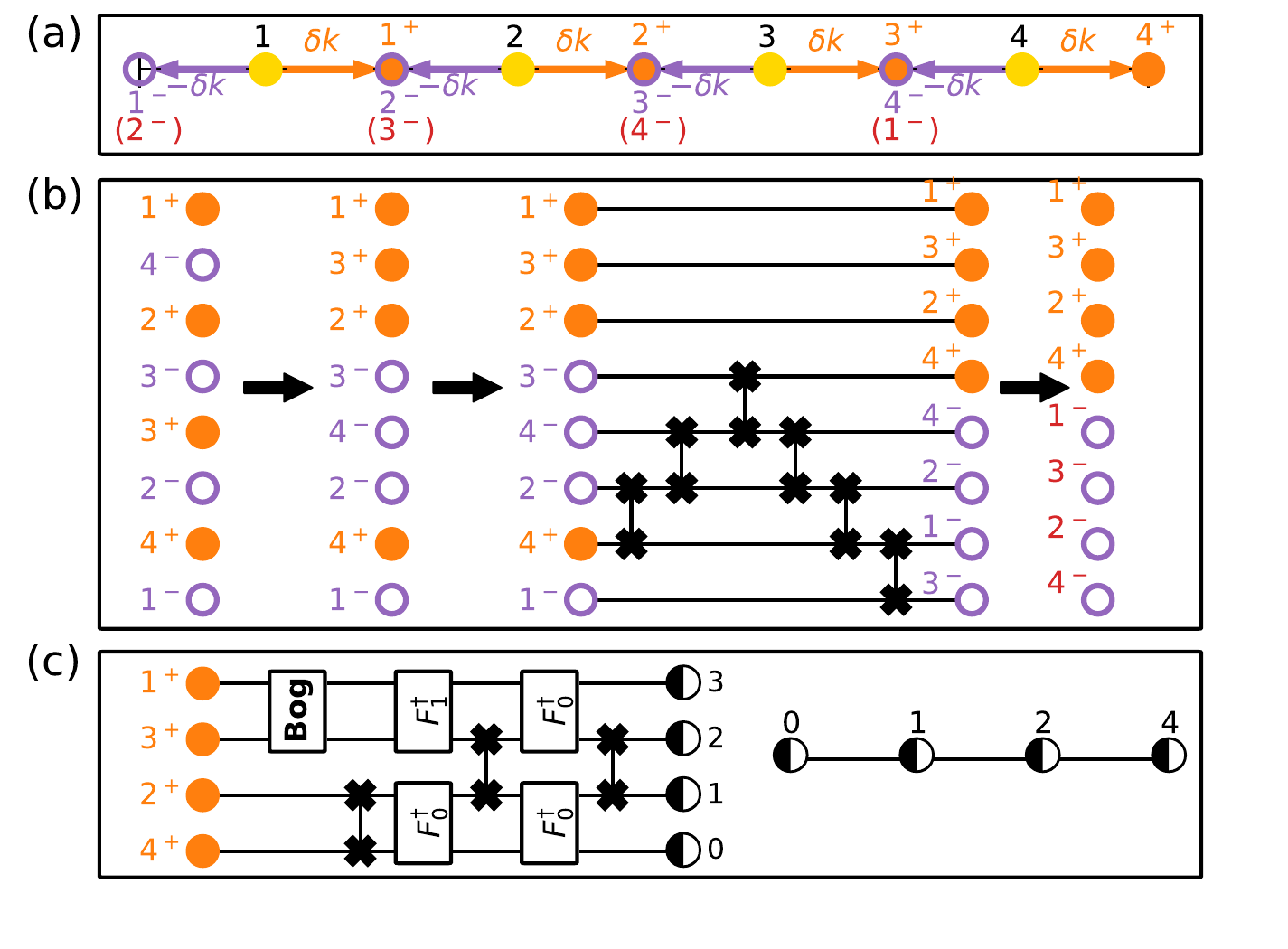}
  \caption{Schematic illustration of the symmetry-enforced circuit using special momentum shifts: (a) momentum points before (the yellow dots) and after shifting (the orange and purple dots); (b) the simplification steps; (c) the quantum circuit transforming one of the copies from the momentum space to position space.}
   \label{figS6}
\end{figure}

Similar procedures can be used to find the entanglement entropy of the bond fermions of the Kitaev honeycomb model. How the entanglement entropy changes with the increasing system size is shown in Fig.~\ref{figS5}. It turns out that an accurate determination of the quantum phase transition requires a system size larger than $L_x\times L_y = 64\times 64$.

\section{Circuit simplification with special momentum shifts} \label{appendix f}

\begin{figure}[htpb]
  \centering
  \includegraphics[width=0.6\columnwidth]{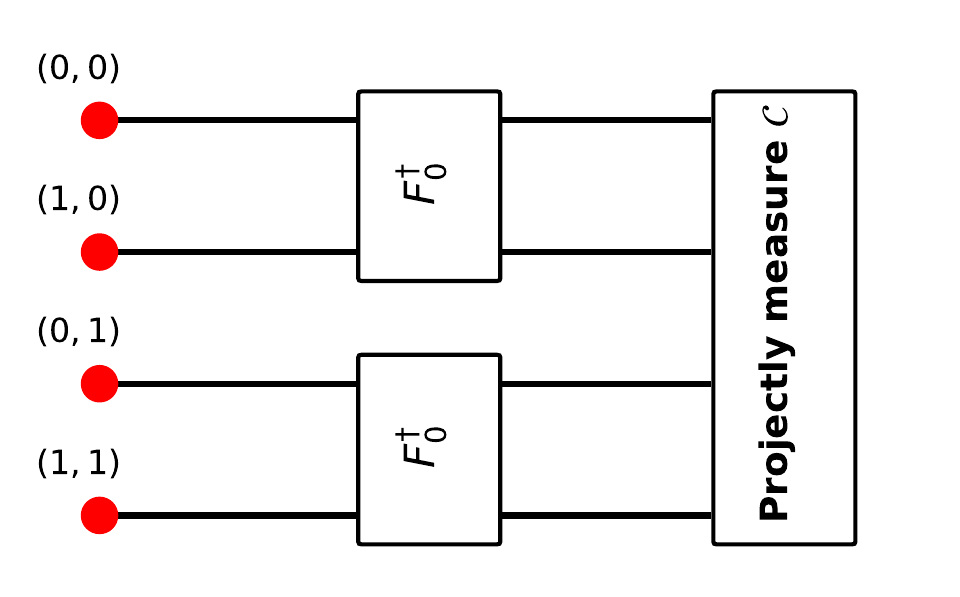}
  \caption{Quantum circuit for measuring entanglement entropy of the Kitaev honeycomb model with the special momentum shift. The labels $(k_x,k_y)$ in the unit of $\pi$ denote the momentum points in the Brillouin zone.}
   \label{figS7}
\end{figure}

Though in general, the symmetry-enforced circuit cannot be simplified, we show here that the circuit can be simplified for special momentum shifts. These special momentum shifts include the high symmetry points.

For a system containg $N$ unit cells, the discrete Fourier transformation and the particle-hole symmetry constraint requires the momentum points to be in the following set:
\beq
\mathcal{K} = \left\{ -\frac{N}{2} + \frac{1}{2}, -\frac{N}{2} + \frac{3}{2}, \cdots , \frac{N}{2} - \frac{1}{2} \right\} \frac{2\pi}{N},
\eeq
where $N=2^n$, with $n$ a positive integer. As demonstrated in the main text, if we shift the momentum points by $\delta k$, their particle-hole partners must shift by $- \delta k$ to maintain particle-hole symmetry. Based on this, we note that $\delta k N/2\pi=1/2$ is a special point, which makes the set of the momentum points by shifting $\delta k$ to be identical to those that shift by $-\delta k$. Therefore, if the gap is \emph{vanishing} at the high-symmetry points, the symmetry-enforced circuit can be decomposed into two identical copies. 

The case with $N=4$ is shown in Fig.~\ref{figS6}. The simplification requires three steps, which are shown in Fig.~\ref{figS6}~(b) from left to right. In the first step, we identify the momentum points labeled by $3^+$ and $4^-$, which is obvious from Fig.~\ref{figS6}~(a); in the second step, due to the gap vanishing at the high-symmetry points, the two-qubit Bogoliubov transformation becomes two separate single-qubit identity operations, so we can change the position of the momentum points $4^+$ and $3^-$ at the cost of a trivial global phase; in the last step, we just renumber the $-\delta k$-shifted momentum points by using the numbers in red shown below the original labels (in purple) in Fig.~\ref{figS6}~(a). Then one can easily find that the result yields two identical copies. We emphasize that the \emph{vanishing} gap at the high-symmetry points is essential in making this simplification possible.

It turns out that the Kitaev honeycomb model fulfills this requirement, so  the simplified circuit can be used to determine the different phases of the model. The details of the simplified circuits are shown in Fig.~\ref{figS7}.


\begin{thebibliography}{f11}
\bibitem{Schlosshauer_RMP_2005} M. Schlosshauer, Decoherence, the measurement problem, and interpretations of quantum mechanics. \href{https://doi.org/10.1103/RevModPhys.76.1267}{Rev. Mod. Phys. \textbf{76}, 1267 (2005)}.
\bibitem{Steane_PRL_1996} A. M. Steane, Error correcting codes in quantum theory. \href{https://doi.org/10.1103/PhysRevLett.77.793}{Phys. Rev. Lett. \textbf{77}, 793 (1996)}.
\bibitem{Knill_arXiv_1996} E. Knill, R. Laflamme, and W. Zurek, Threshold accuracy for quantum computation. \href{https://arxiv.org/abs/quant-ph/9610011}{arXiv:9610011 [quant-ph] (1996)}.
\bibitem{Aharonov_ACM_1997} D. Aharonov, and M. Ben-Or, Fault-tolerant quantum computation with constant error. \href{https://doi.org/10.1137/S0097539799359385}{SIAM J. Comput., 38, 1207 (2008)}.
\bibitem{Terhal_RMP_2015} B. M. Terhal, Quantum error correction for quantum memories. \href{https://doi.org/10.1103/RevModPhys.87.307}{Rev. Mod. Phys. \textbf{87}, 307 (2015)}.
\bibitem{Aliferis_QIC_2005} A. Aliferis, D. Gottesman, and, J. Preskill, Quantum accuracy threshold for concatenated distance-$3$ codes \href{https://dl.acm.org/doi/10.5555/2011665.2011666}{Quantum Inf. Comput. \textbf{6}, 97 (2005)}.
\bibitem{Bacon_PRA_2006} D. Bacon, Operator quantum error-correcting subsystems for self-correcting quantum memories. \href{https://doi.org/10.1103/PhysRevA.73.012340}{Phys. Rev. A \textbf{73}, 012340 (2006)}.
\bibitem{Fowler_PRA_2012} A. G. Fowler, M. Mariantoni, J. M. Martinis, and A. N. Cleland, Surface codes: towards practical large-scale quantum computation. \href{https://doi.org/10.1103/PhysRevA.86.032324}{Phys. Rev. A \textbf{86}, 032324 (2012)}.
\bibitem{Tuckett_PRL_2018} D. K. Tuckett, S. D. Bartlett, and S. T. Flammia, Ultrahigh error threshold for surface codes with biased noise. \href{https://doi.org/10.1103/PhysRevLett.120.050505}{Phys. Rev. Lett. \textbf{120}, 050505 (2018)}.
\bibitem{Chao_prl_2018} R. Chao, and B. W. Reichardt, Quantum error correction with only two extra qubits. \href{https://doi.org/10.1103/PhysRevLett.121.050502}{Phys. Rev. Lett. \textbf{121}, 050502 (2018)}.
\bibitem{Li_PRX_2019} M. Li, D. Miller,  M. Newman, Y. Wu, and K. R. Brown, $2D$ Compass Codes. \href{https://doi.org/10.1103/PhysRevX.9.021041}{Phys. Rev. X \textbf{9}, 021041 (2019)}.
\bibitem{Kitaev_AP_1997} A. Kitaev, Fault-tolerant quantum computation by anyons. \href{https://doi.org/10.1016/S0003-4916(02)00018-0}{Ann. Phys. \textbf{303}, 2 (2003)}.
\bibitem{Nayak_RMP_2008} C. Nayak, S. H. Simon, A. Stern, M. Freedman, and S. D. Sarma, Non-Abelian anyons and topological quantum computation. \href{https://doi.org/10.1103/RevModPhys.80.1083}{Rev. Mod. Phys. \textbf{80}, 1083 (2008)}.
\bibitem{Sarma_NPJ_2015} S. D. Sarma, M. Freedman, and C. Nayak, Majorana zero modes and topological quantum computation. \href{https://doi.org/10.1038/npjqi.2015.1}{NPJ Quantum Information \textbf{1}, 15001 (2015)}.
\bibitem{Kitaev_AP_2006} A. Kitaev, Anyons in an exactly solved model and beyond. \href{https://doi.org/10.1016/j.aop.2005.10.005}{Ann. Phys. \textbf{321}, 2 (2006)}.
\bibitem{Yao_PRL_2007} H. Yao, and S. A. Kivelson, Exact chiral spin liquid with Non-Abelian anyons. \href{https://doi.org/10.1103/PhysRevLett.99.247203}{Phys. Rev. Lett. \textbf{99}, 247203 (2007)}.
\bibitem{Yang_PRB_2007} S. Yang, D. L. Zhou, and C. P. Sun, Mosaic spin models with topological order. \href{https://doi.org/10.1103/PhysRevB.76.180404}{Phys. Rev. B \textbf{76}, 180404 (2007)}.
\bibitem{Feng_PRL_2007} X.-Y. Feng, G. M. Zhang, and T. Xiang, Topological characterization of quantum phase Transitions in a Spin-$1/2$ Model. \href{https://doi.org/10.1103/PhysRevLett.98.087204}{Phys. Rev. Lett. \textbf{98}, 087204 (2007)}.
\bibitem{Lee_PRL_2007} D. H. Lee, G. M. Zhang, and T. Xiang, Edge solitons of topological insulators and fractionalized quasiparticles in Two Dimensions. \href{https://doi.org/10.1103/PhysRevLett.99.196805}{Phys. Rev. Lett. \textbf{99}, 196805 (2007)}.
\bibitem{Si_arXiv_2007} T. Si, and Y. Yu, Exactly soluble spin-$1/2$ models on three-dimensional lattices and non-abelian statistics of closed string excitations. \href{https://arxiv.org/abs/0709.1302}{arXiv:0709.1302 [cond-mat] (2007)} .
\bibitem{Yu_EPL_2008} Y. Yu, \& Z. Q. Wang, An exactly soluble model with tunable p-wave paired fermion ground states. \href{https://doi.org/10.1209/0295-5075/84/57002}{Europhys. Lett. \textbf{84}, 57002 (2008)}.
\bibitem{Baskaran_arXiv_2009} G. Baskaran, G. Santhosh, and R. Shankar, Exact quantum spin liquids with Fermi surfaces in spin-half models. \href{https://arxiv.org/abs/0908.1614}{arXiv:0908.1614 [cond-mat] (2009)}.
\bibitem{Mandal_PRB_2009} S. Mandal, and N. Surendran, Exactly solvable Kitaev model in three dimensions. \href{https://doi.org/10.1103/PhysRevB.79.024426}{Phys. Rev. B \textbf{79}, 024426 (2009)}.
\bibitem{Ryu_PRB_2009} S. Ryu, Three-dimensional topological phase on the diamond lattice. \href{https://doi.org/10.1103/PhysRevB.79.075124}{Phys. Rev. B \textbf{79}, 075124 (2009)}.
\bibitem{Yao_PRL_2009} H. Yao, S.-C. Zhang, and S. A. Kivelson, Algebraic spin liquid in an exactly solvable spin model. \href{https://doi.org/10.1103/PhysRevLett.102.217202}{Phys. Rev. Lett. \textbf{102}, 217202 (2009)}.
\bibitem{Wu_PRB_2009} C. Wu, D. Arovas, and H. H. Hung, $\Gamma$-matrix generalization of the Kitaev model. \href{https://doi.org/10.1103/PhysRevB.79.134427}{Phys. Rev. B \textbf{79}, 134427 (2009)}.
\bibitem{Tikhonov_PRL_2010} K. S. Tikhonov, and M. V. Feigelman, Quantum spin metal state on a decorated honeycomb lattice. \href{https://doi.org/10.1103/PhysRevLett.105.067207}{Phys. Rev. Lett. \textbf{105}, 067207 (2010)}.
\bibitem{Chern_PRB_2010} G. W. Chern, Three-dimensional topological phases in a layered honeycomb spin-orbital model. \href{https://doi.org/10.1103/PhysRevB.81.125134}{Phys. Rev. B \textbf{81}, 125134 (2010)}.
\bibitem{Wang_PRB_2010} F. Wang, Realization of the exactly solvable Kitaev honeycomb lattice model in a spin-rotation-invariant system. \href{https://doi.org/10.1103/PhysRevB.81.184416}{Phys. Rev. B \textbf{81}, 184416 (2010)}.
\bibitem{Lahtinen_PRB_2010} V. Lahtinen, and J. K. Pachos, Topological phase transitions driven by gauge fields in an exactly solvable model. \href{https://doi.org/10.1103/PhysRevB.81.245132}{Phys. Rev. B \textbf{81}, 245132 (2010)}.
\bibitem{Kells_NJP_2011} G. Kells, J. Kailasvuori, J. K. Slingerland, and J. Vala, Kaleidoscope of topological phases with multiple Majorana species. \href{https://doi.org/10.1088/1367-2630/13/9/095014}{New J. Phys. \textbf{13}, 095014 (2011)}.
\bibitem{Yao_PRL_2011} H. Yao, and D. H. Lee,  Fermionic magnons, Non-Abelian spinons, and the spin quantum hall effect from an exactly solvable spin-$1/2$ Kitaev model with $SU(2)$ symmetry. \href{https://doi.org/10.1103/PhysRevLett.107.087205}{Phys. Rev. Lett. \textbf{107}, 087205 (2011)}.
\bibitem{Lai_PRB_2011} H. H. Lai, and O. I. Motrunich, Power-law behavior of bond energy correlators in a Kitaev-type model with a stable parton Fermi surface. \href{https://doi.org/10.1103/PhysRevB.83.155104}{Phys. Rev. B \textbf{83}, 155104 (2011)}.
\bibitem{Chua_PRB_2011} V. Chua, H. Yao, \& G. A. Fiete, Exact chiral spin liquid with stable spin Fermi surface on the kagome lattice. \href{https://doi.org/10.1103/PhysRevB.83.180412}{Phys. Rev. B \textbf{83}, 180412 (2011)}.
\bibitem{Nakai_PRB_2012} R. Nakai, S. Ryu, and A. Furusaki, Time-reversal symmetric Kitaev model and topological superconductor in two dimensions. \href{https://doi.org/10.1103/PhysRevB.85.155119}{Phys. Rev. B \textbf{85}, 155119 (2012)}.
\bibitem{Nussinov_RMP_2015} Z. Nussinov, and J. van den Brink, Compass and Kitaev models: theory and physical motivations. \href{https://doi.org/10.1103/RevModPhys.87.1}{Rev. Mod. Phys. \textbf{87}, 1 (2015)}.
\bibitem{Hermanns_PRL_2015} M. Hermanns, K. O'Brien, and S. Trebst, Weyl spin liquids. \href{https://doi.org/10.1103/PhysRevLett.114.157202}{Phys. Rev. Lett. \textbf{114}, 157202 (2015)}.
\bibitem{O'Brien_PRB_2016} K. O'Brien, M. Hermanns, \& S. Trebst, Classification of gapless $Z_2$ spin liquids in three-dimensional Kitaev models. \href{https://doi.org/10.1103/PhysRevB.93.085101}{Phys. Rev. B \textbf{93}, 085101 (2016)}.
\bibitem{Chen_PRL_2018} Z. Chen, X. Li, and T. K. Ng, Exactly Solvable BCS-Hubbard Model in Arbitrary Dimensions. \href{https://doi.org/10.1103/PhysRevLett.120.046401}{Phys. Rev. Lett. \textbf{120}, 046401 (2018)}.
\bibitem{Miao_arXiv_2018} J.-J. Miao, H.-K. Jin, F.-C. Zhang, and Y. Zhou, Exact solution to a class of generalized Kitaev spin-1/2 models in arbitrary dimensions. \href{https://doi.org/10.1007/s11433-019-1442-2}{Sci. China Phys. Mech. Astron. \textbf{63}, 247011 (2020)}.
\bibitem{Miao_PRB_2018} J.-J. Miao, H.-K. Jin, F. Wang, F.-C. Zhang, and Y. Zhou, Pristine Mott insulator from an exactly solvable spin-$1/2$ Kitaev model. \href{https://doi.org/10.1103/PhysRevB.99.155105}{Phys. Rev. B \textbf{99}, 155105 (2019)}.
\bibitem{Stern_Science_2013} A. Stern, and N. H. Lindner, Topological quantum computation—from basic concepts to first experiments. \href{https://doi.org/10.1126/science.1231473}{Science \textbf{339}, 1179-1184 (2013)}.
\bibitem{Yao_PRL_2010} H. Yao, and X.-L. Qi, Entanglement entropy and entanglement spectrum of the Kitaev model. \href{https://doi.org/10.1103/PhysRevLett.105.080501}{Phys. Rev. Lett. \textbf{105}, 080501 (2010)}.
\bibitem{Meichanetzidis_PRB_2016} K. Meichanetzidis, M. Cirio, J. K. Pachos and V. Lahtinen, Anatomy of fermionic entanglement and criticality in Kitaev spin liquids. \href{https://doi.org/10.1103/PhysRevB.94.115158}{Phys. Rev. B \textbf{94}, 115158 (2016)}.
\bibitem{Schmoll_PRB_2017} P. Schmoll, and R. Orus, Kitaev honeycomb tensor networks: Exact unitary circuits and applications. \href{https://doi.org/10.1103/PhysRevB.95.045112}{Phys. Rev. B \textbf{95}, 045112 (2017)}.
\bibitem{Verstraete_PRA_2009} F. Verstraete, J. I. Cirac,, and J. I. Latorre, Quantum circuits for strongly correlated quantum systems. \href{https://doi.org/10.1103/PhysRevA.79.032316}{Phys. Rev. A \textbf{79}, 032316 (2009)}.
\bibitem{Lierta_Qu_2018} A. Cervera-Lierta, Exact Ising model simulation on a quantum computer. \href{https://doi.org/10.22331/q-2018-12-21-114}{Quantum \textbf{2}, 114 (2018)}.
\bibitem{Ferris_PRL_2014} A. J. Ferris, Fourier transform for fermionic systems and the spectral tensor network. \href{https://doi.org/10.1103/PhysRevLett.113.010401}{Phys. Rev. Lett. \textbf{113}, 010401 (2014)}.
\bibitem{Ivanov_PRL_2001} D. A. Ivanov, Non-Abelian Statistics of Half-Quantum Vortices in $p$-Wave Superconductors. \href{https://doi.org/10.1103/PhysRevLett.86.268}{Phys. Rev. Lett. \textbf{86}, 268 (2001)}.
\bibitem{Vidal_PRA_2004} G. Vidal, and C. M. Dawson, Universal quantum circuit for two-qubit transformations with three controlled-NOT gates. \href{https://doi.org/10.1103/PhysRevA.69.010301}{Phys. Rev. A \textbf{69}, 010301 (2004)}.
\bibitem{Smolin_PRL_2012} J. A. Smolin, J. M. Gambetta, and G. Smith, Efficient method for computing the maximum-likelihood quantum state from measurements with additive gaussian noise. \href{https://doi.org/10.1103/PhysRevLett.108.070502}{Phys. Rev. Lett. \textbf{108}, 070502 (2012)}.
\bibitem{qiskit2019} G. Aleksandrowicz, {\it et al.} Qiskit: An open-source framework for quantum computing. \href{https://doi.org/10.5281/ZENODO.2562111 }{https://doi.org/10.5281/ZENODO.2562111} (2019).
\bibitem{Peschel_JPA_2003} I. Peschel, Calculation of reduced density matrices from correlation functions. \href{https://iopscience.iop.org/article/10.1088/0305-4470/36/14/101}{J. Phys. A: Math.Gen. \textbf{36}, L205 (2003)}.
\bibitem{Vidal_PRL_2003} G. Vidal, J. I. Latorre, E. Rico, and A. Kitaev, Entanglement in Quantum Critical Phenomena. \href{https://doi.org/10.1103/PhysRevLett.90.227902}{Phys. Rev. Lett. \textbf{90}, 227902 (2003)}.
\bibitem{Jiang_arXiv_2018} H. Jiang, C.-Y. Wang, B. Huang, and Y.-M. Lu, Field induced quantum spin liquid with spinon Fermi surfaces in the Kitaev model. \href{https://arxiv.org/abs/1809.08247}{arXiv:1809.08247 [cond-mat] (2018)} .
\bibitem{Schnyder_PRB_2008} A. P. Schnyder, S. Ryu, A. Furusaki, and A. W. W. Ludwig, Classification of topological insulators and superconductors in three spatial dimensions. \href{https://doi.org/10.1103/PhysRevB.78.195125}{Phys. Rev. B \textbf{78}, 195125 (2008)}.
\end{thebibliography}
\end{document}